\documentstyle[epsfig,emulateapj]{article}

\received{ }
\accepted{ }

\lefthead{Catanese \& Weekes}
\righthead{VHE Gamma-Ray Astronomy}

\submitted{Invited Review to appear in Publ. Astron. Soc. of the Pacific}

\begin{document}

\title{Very High Energy Gamma-Ray Astronomy}

\author{Michael Catanese}
\affil{Department of Physics and Astronomy, Iowa State
University, Ames, IA 50011-3160 \\
Electronic mail: mcatanese@cfa.harvard.edu}
\centerline{and}
\author{Trevor C. Weekes}
\affil{ Fred Lawrence Whipple Observatory, Harvard-Smithsonian 
CfA, P.O. Box 97, Amado, AZ 85645-0097 \\ 
Electronic mail: tweekes@cfa.harvard.edu}

\authoremail{tweekes@cfa.harvard.edu}
\authoraddr{Trevor Weekes, F. L. Whipple Observatory, 670 Mt. Hopkins Road,
Amado, AZ 85645}

\begin{abstract} 

We present a review of the current status of very high energy
$\gamma$-ray astronomy. The development of the atmospheric Cherenkov
imaging technique for ground-based $\gamma$-ray astronomy has led to a
rapid growth in the number of observatories. The detection of TeV
$\gamma$-rays from Active Galactic Nuclei was unexpected and is
providing new insights into the emission mechanisms in the jets.  Next
generation telescopes are under construction and will increase
dramatically the knowledge available at this extreme end of the cosmic
electromagnetic spectrum.

\end{abstract}


\setcounter{footnote}{0}
\section{Introduction}
\label{intro}

Very high energy (VHE) $\gamma$-ray astronomy, (defined here as
observations at energies above 300\,GeV and below 100\,TeV) became
viable with the development of the atmospheric Cherenkov imaging
technique which crossed the vital detection threshold with the
detection of the Crab Nebula ten years ago (\cite{weekes89}). Although
at the periphery of the observable electromagnetic spectrum, the VHE
band must now be considered a legitimate astronomical discipline with
established sources, both steady and variable, both galactic and
extragalactic, a growing number of observatories, and the promise of
significant advances in detection techniques in the next few
years. The confirmation of the detection of the Crab Nebula by more
than eight groups over the past decade has given credibility to the
existence of sources of TeV $\gamma$-rays and led to a rapid
improvement in the sensitivity of ground-based $\gamma$-ray detection
techniques.

The challenge therefore in this short review, which is aimed at the
general astronomical community, is to convince the reader that there
are viable methods of detecting $\gamma$-rays of energy 300\,GeV and
above from the ground, that a population of credible sources of
various classes have been detected, and that these detections make a
significant contribution to the astrophysics of high energy
sources. Although, as we shall see later, there are plans to extend
the ground-based detection techniques down to energies of 20\,GeV, and
there are viable air shower experiments that operate at energies of 50\,TeV 
and above, this review will focus on observations in the 300\,GeV
to 30\,TeV energy range; this has been the most studied energy band
because it is easily accessible to the atmospheric Cherenkov imaging
technique.

\section{Techniques and Instrumentation}
\label{technical}

\subsection{Space telescopes}
\label{space-tels}

The physics involved in the interaction of photons with matter is well
established.  At energies above 10\,MeV the predominant interaction is
pair production in which a $\gamma$-ray  is converted into an
electron-positron pair in the presence of a nucleus: E$_\gamma 
\rightarrow $
m$_{e^+}$c$^2$ + m$_{e^-}$c$^2$.  The electron and positron carry
information about the direction, energy and polarization of the
primary $\gamma$-ray and hence detection methods focus on the
observation
of these secondary particles.  The interaction length of
photons in matter is 30\,g-cm$^2$ and the earth's atmosphere is
1030\,g-cm$^2$ thick, so the preferred (only) method of unambiguously
detecting high energy $\gamma$-rays is from space vehicles (from
balloons initially but now from satellites).

The basic elements of a space-borne $\gamma$-ray detector are
therefore (1) a particle detector in which the $\gamma$-ray interacts
and the resulting electron pair tracks are recorded; (2) a calorimeter
in which the electrons are absorbed and
their total energy recorded; (3) an anti-coincidence shield
surrounding the tracking detector which registers (and rejects) the
incidence of charged cosmic-ray particles (which are about 10,000
times more numerous than the $\gamma$-rays). In most of the telescopes
flown to date, the tracking detector has been a spark chamber, the
calorimeter a Sodium Iodide crystal and the anti-coincidence detector
a thin sheet of scintillator. The resulting telescope is a
sophisticated device which can unambiguously identify $\gamma$-rays,
has an energy resolution of about 15\%, an angular resolution of
1$^\circ$ and a field of view of 20$^\circ$ to 40$^\circ$ half
angle. Unfortunately this sophistication does not come cheaply; the
effective collection area is only a small fraction of the total
telescope area so that in the largest telescope flown to date, the
hugely successful Energetic Gamma Ray Experiment Telescope (EGRET) on
the {\it Compton Gamma Ray Observatory} ({\it CGRO})
(\cite{Thompson95}; \cite{Hartman99}), the effective collection area
was only 1,500\,cm$^2$ (about the size of two pages of this journal!)
whereas the actual physical instrument was about the size of a compact
car. All of the $\gamma$-ray telescopes flown to date (the SAS-II
telescope in 1973, the COS-B telescope in 1975, EGRET in 1991) 
have had the same functional form;
the next generation space telescope, the Gamma-ray Large Area Space
Telescope (GLAST), scheduled for launch in 2005 (\cite{gehrels99}) has
the same general features but will use solid state detectors with a
factor of 10-30 improvement in sensitivity.

The Third EGRET catalog (based on some four years of observation by
EGRET) (\cite{Hartman99}) contains a listing of more than 250 sources
of $>$100\,MeV $\gamma$-rays, more than half of which are unidentified
with any known astronomical object.  In addition to a detailed map of
the diffuse emission along the galactic plane, there is evidence for
more than 100 galactic sources, a small number of which have been
identified with pulsars and possibly with supernova remnants.  The
bulk of the sources away from the plane are extragalactic and have
been identified with active galactic nuclei (AGNs), almost all of
which are blazars. Some of the sources are variable indicating that
the high energy $\gamma$-ray sky is a dynamic place. Many of the
identified sources (pulsars, AGNs) have very flat spectra and have
luminosities that peak in the high energy region of the spectrum. The
EGRET sensitivity extends to 10 GeV but is limited by the calorimeter
(at the highest energies the cascade from the electron-positron pair
is not contained and charged particles from the cascade can reach the
anti-coincidence detector, causing a veto of the event). Since the
source flux almost invariably decreases as energy increases, it is
only possible to extend observations by building larger
telescopes. Even for a flat spectrum source (power law with
differential spectral index -2.0), the power sensitivity of EGRET
falls off with energy.

\subsection{Ground-based Telescopes}

The earth's atmosphere is as opaque to photons of energy $>$300\,GeV
as it is to photons in the {\it CGRO} range (100\,keV to 30\,GeV).
However, at these higher energies the effects of atmospheric
absorption are detectable at ground level, either as a shower of
secondary particles from the resulting electromagnetic cascade or as a
flash of Cherenkov light from the passage of these particles through
the earth's atmosphere. As in the {\it CGRO} range, $\gamma$-ray
observations are severely limited by the charged cosmic particle flux
which gives superficially similar signals at ground-level and, for a
given photon energy, is 10,000 times more numerous. It is not possible
to veto out the charged cosmic-ray background with an anti-coincidence
shield.  As such, it might seem impossible to do $\gamma$-ray
astronomy with such indirect techniques. However there are small, but
significant, differences in the cascades resulting from the impact of
a photon and a proton on the upper atmosphere; also the
electromagnetic cascade retains the original direction of the photon
to a high degree, and the spread of secondary particles and Cherenkov
photons is so large so that a simple detector can have an incredible
(by space-based $\gamma$-ray detector standards) collection area.

\begin{figure*}[t]
\centerline{\epsfig{file=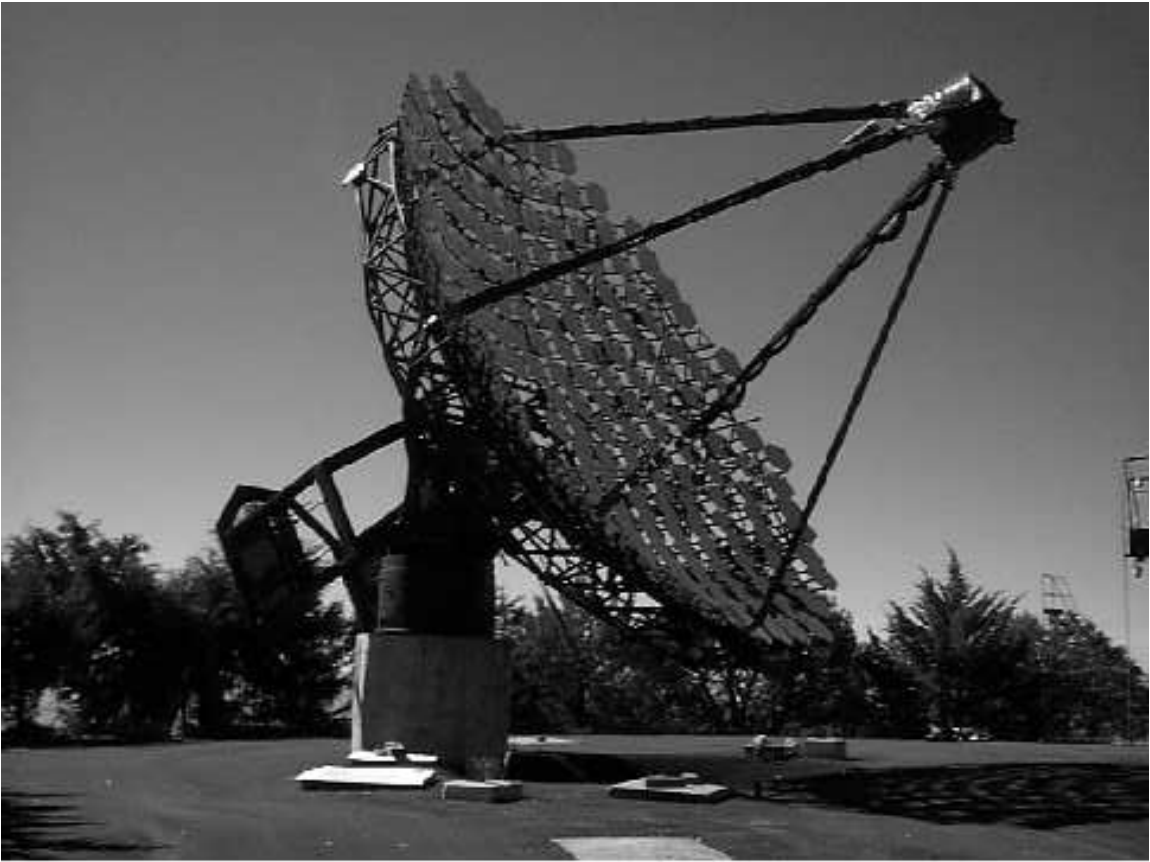,width=4.5in,angle=270.}}
\caption{The Whipple Observatory 10-m imaging atmospheric Cherenkov
telescope.  
\label{10m-fig}}
\end{figure*}

In practice the most successful detectors are atmospheric Cherenkov
imaging telescopes (ACITs) which record the images of the Cherenkov
light flashes and which can identify the images of electromagnetic
cascades from putative sources with 99.7\% efficiency
(\cite{aharaker98}; \cite{ong98}).  Originally proposed in 1977
(\cite{weetur77}), the technique was not demonstrated until ten years
later when the Whipple Observatory 10-m reflector
(Figure~\ref{10m-fig}) was equipped with a primitive imaging camera
and used to detect the Crab Nebula (\cite{weekes89}). The technology
was not new (arrays of fast photomultiplier tubes in the focal plane
of large optical reflectors with readout through standard fast
amplifiers, discriminators and analog-to-digital converters) but the
technique was only fully exploited in the past decade. Compared to
high energy space telescopes such as EGRET, ACITs have large
collection areas ($>$50,000m$^2$) and high angular resolution
($\sim$0.1$^\circ$).  ACITs also have reasonably good energy
resolution ($\sim20-40\%$), but small fields of view (FOV)
($<$5$^{\circ}$) and a background of diffuse cosmic electrons which
produce electromagnetic cascades identical to those of $\gamma$-rays.
ACITs also have low duty cycles ($<$10\%) because the Cherenkov
signals are faint and produced at altitudes of several kilometers,
requiring cloudless, moonless skies for observations.  Most of the
results reported to date have been in the energy range 300\,GeV to
30\,TeV.

In recent years VHE $\gamma$-ray astronomy has seen two major
advances: first, the development of high resolution ACITs has
permitted the efficient rejection of the hadronic background, and
second, the construction of arrays of ACITs has improved the 
measurement of the energy spectra from $\gamma$-ray sources. 
The first is exemplified by the Whipple Observatory 10-m telescope
with more modern versions, CAT, a French telescope in Pyren\'ees
(\cite{barrau98}), and CANGAROO, a Japanese-Australian telescope in
Woomera, Australia (\cite{hara93}). The most significant examples of
the second are HEGRA, a five telescope array of small imaging
telescopes on La Palma in the Canary Islands run by an
Armenian-German-Spanish collaboration (\cite{daum97}), and the Seven
Telescope Array in Utah, which is operated by a group of Japanese
institutions (\cite{Aiso97}). These techniques are relatively mature
and the results from contemporaneous observations of the same source
with different telescopes are consistent (\cite{Protheroe97}).
Vigorous observing programs are now in place at all of these
facilities.  A vital observing threshold has been achieved whereby
both galactic and extragalactic sources have been reliably
detected. Many exciting results are anticipated as more of the sky is
observed with this present generation of telescopes.

The atmospheric Cherenkov imaging technique has now been adopted at a
number of observatories whose properties are summarized in
Table~\ref{tels-table}.

\begin{table*}
\begin{center}
\caption{Operating ACIT Observatories c. 1999 May \label{tels-table}}
{\footnotesize
\begin{tabular}{ccccccl} \hline\hline
Group & Countries & Location & Telescope(s) & Camera & Threshold & 
\multicolumn{1}{c}{Epoch} \\
 &  &  & Number$\times$Aperture & Pixels & (TeV) & 
\multicolumn{1}{c}{Beginning} \\ \hline
Whipple & USA-UK-Irel.& Arizona,USA & 10\,m & 331 & 250 & 1984 \\
Crimea  & Ukraine & Crimea & 6$\times$2.4\,m & 6$\times$37 & 1 & 1985 \\
SHALON & Russia & Tien Shen, Russia & 4\,m & 244 & 1.0 & 1994 \\
CANGAROO & Japan-Aust. & Woomera,Aust. & 3.8\,m & 256 & 0.5 & 1994 \\
HEGRA & German-Armen.-Sp. & La Palma, Sp. & 5$\times$3\,m & 5$\times$271  & 0.5 & 1994 \\
CAT & France & Pyren\'ees & 3\,m & 600 & 0.25 & 1996 \\
Durham & UK & Narrabri,Aust. & 3$\times$7\,m & 1$\times$109 & 0.25 & 1996 \\
TACTIC & India & Mt.Abu,India & 10\,m & 349 & 0.3 & 1997 \\
Seven TA & Japan & Utah,USA & 7$\times$2\,m & 7$\times$256 & 0.5 & 1998 \\ \hline
\end{tabular}
}
\end{center}
\end{table*}

Above 30\,TeV there are enough residual particles in the
electromagnetic cascades that they can be detected at high mountain
altitudes using arrays of particle detectors and fast wavefront timing
(\cite{ong98}). These arrays have large collection areas
($>$10,000\,m$^2$), good angular resolution ($\sim$0.5$^\circ$),
moderate energy resolution ($\sim100\%$), good duty cycle (100\%) and
large FOV ($\sim$2\,sr); however their ability to discriminate
$\gamma$-rays from charged cosmic rays is severely limited. Despite
the early promise of these experiments, which led to a considerable
investment in their construction and operation, no verifiable
detections have been reported by particle air shower arrays.

\section{Galactic Sources}
\label{gal-sec}

\subsection{The Crab Nebula}
\label{crab-sec}

The Crab Nebula was first detected with a 37 pixel camera on the
Whipple Observatory 10-m optical reflector in 1989.  This early and
somewhat crude instrument yielded a 9\,$\sigma$ detection with some 60
hours of integration on the source (\cite{weekes89}). The detection
relied on discrimination of the $\gamma$-ray images from the much more
numerous hadron background images. The possibility of a systematic
effect with a new, and not yet proven, technique could not be
completely discounted (although numerous tests were made for
consistency).  Nonetheless it required the independent confirmation of
the detection by other groups using different versions of the
technique to really convince skeptics. The Whipple group subsequently
detected the source at the 20\,$\sigma$ level using an upgraded camera
(109 pixels) (\cite{vacanti91}) and now routinely detects the source
at the 5-6\,$\sigma$ level in an hour of observation. The detected
photon rate (about 2 per minute) is more than that registered by EGRET
at its optimum energy (100 MeV).

Since then the Crab has been detected by eight independent groups
using different versions of the atmospheric Cherenkov imaging
technique (including one group in the Southern Hemisphere). Some of
these detections are listed in Table~\ref{crab-table}.  The energy
spectrum is now well determined at energies between 300\,GeV and
50\,TeV (\cite{Hillas98};\cite{Tanimori98b}). To date there
have been no positive detections reported by air shower array
experiments using particle detectors (which operate at somewhat higher
energies)(\cite{ong98}).

\begin{table*}[t]
\caption{Flux from the Crab Nebula \label{crab-table}}
\begin{center}
\begin{tabular}{cll} \hline\hline
Group & \multicolumn{1}{c}{VHE Spectrum} & \multicolumn{1}{c}{E$_{\rm th}$} \\ 
 & \multicolumn{1}{c}{($10^{-11}$ photons cm$^{-2}$ s$^{-1}$)} & 
\multicolumn{1}{c}{(TeV)} \\ \hline
Whipple (1991)$^a$ & ${\rm (25 (E/0.4TeV))^{-2.4 \pm 0.3}}$  & 0.4 \\
Whipple (1998)$^b$ & ${\rm (3.2 \pm 0.7) (E/TeV)^{(-2.49 \pm 0.06_{stat} 
 \pm 0.05_{syst}}}$ & 0.3 \\
HEGRA (1999)$^c$ & ${\rm (2.7 \pm 0.2 \pm 0.8) (E/TeV)^{-2.61 \pm 0.06_{stat} 
 \pm 0.10_{syst}}}$ & 0.5 \\
CAT (1998)$^d$ & ${\rm (2.7 \pm 0.17 \pm 0.40) (E/TeV)^{-2.57 \pm 0.14_{stat} 
 \pm 0.08_{syst}}}$ & 0.25 \\ \hline
\multicolumn{3}{l}{$^a$\protect\cite{vacanti91}} \\
\multicolumn{3}{l}{$^b$\protect\cite{Hillas98}} \\
\multicolumn{3}{l}{$^c$ Priv. Com.: A. Konopelko} \\
\multicolumn{3}{l}{$^d$ Priv. Com.: M. Punch} \\
\end{tabular}
\end{center}
\end{table*}

\begin{figure*}
\centerline{\epsfig{file=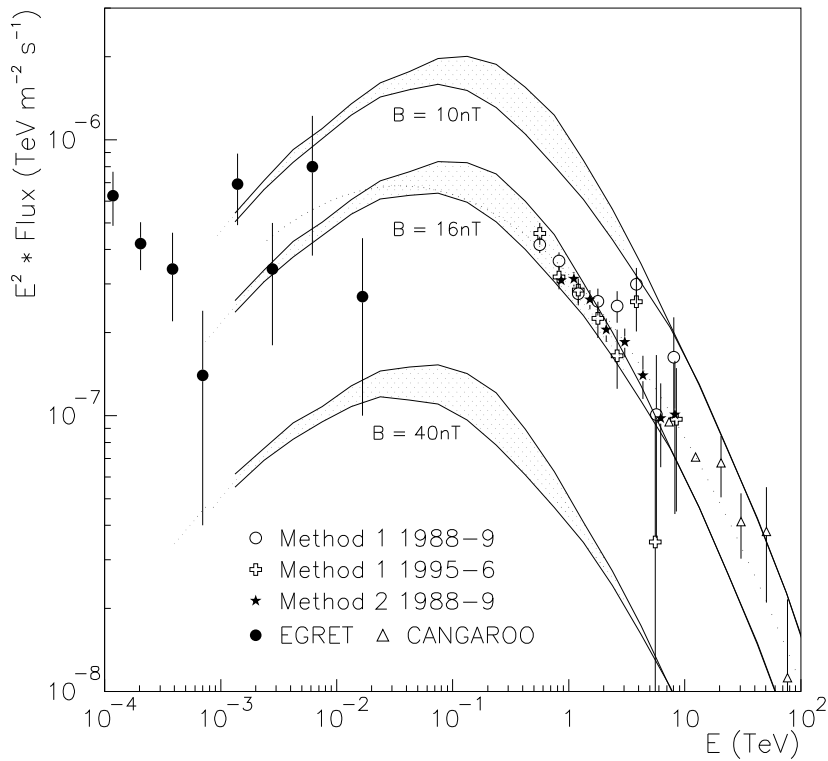,height=4in,angle=0.}}
\caption{The VHE spectral energy distribution of the Crab Nebula
compared with the predictions of a synchrotron self-Compton emission
model (\protect\cite{Hillas98}).
\label{crab-spec}}
\end{figure*}
            
The simple Compton-synchrotron model (\cite{Gould64}) has been updated
to take account of a better understanding of the nebula
(\cite{dejager92}; \cite{Hillas98}); the measured flux is in good
agreement with the predicted flux for a value of magnetic field
($1.6\pm0.1 \times 10^{-4}$\,G) that is slightly lower than the
equipartition value (Figure~\ref{crab-spec}). Although this model is
certainly simplistic given the structure now seen in optical images of
the nebula, it shows that there is a viable mechanism that must work
at some level.

As in many other bands of the electromagnetic spectrum, the Crab
Nebula has become the standard candle for TeV $\gamma$-ray astronomy.
Most importantly perhaps it is available as a steady source to test
and calibrate the ACIT and can be seen from both hemispheres.
Improvements in analysis techniques developed on Crab Nebula data
have led directly to the detections of the AGNs discussed below.

\subsection{Supernova Remnants: Plerions}
\label{pler-sec}

\begin{table*}[t]
\caption{TeV Observations of Plerions \label{plerion-table}}
\begin{center}
\begin{tabular}{cccl} \hline\hline
Source & Energy & Flux/Upper Limit & \multicolumn{1}{c}{Group} \\
 & (GeV) & ($\times$10$^{-11}$\,cm$^{-2}$\,s$^{-1}$) & \\ \hline
Crab Nebula    &  400 & 7.0  & Whipple, ASGAT, HEGRA, TA \\
               &       & & Crimea, *Gamma, CANGAROO, CAT \\
PSR\,1706-44   & 1000 & 0.8  & CANGAROO, Durham  \\
Vela           & 2500  & 0.29  & CANGAROO \\
SS\,433        & 550  &  $<$1.8  & Whipple  \\
3C\,58         & 550  &  $<$1.1  & Whipple  \\
PSR\,0656+14   & 1000 &  $<$3.4  & Whipple  \\ \hline
\end{tabular}
\end{center}
\end{table*}

The Crab Nebula is a somewhat unique object and hence one could not
confidently predict what other supernova remnants might be
detectable. The Crab is a member of that sub-class of supernova
remnants known as plerions in which a bubble of relativistic particles
is powered by a central pulsar. No other plerions have been seen by
telescopes in the Northern Hemisphere (\cite{Reynolds93})
(Table~\ref{plerion-table}).  However two have been detected in the
Southern Hemisphere by the CANGAROO group. They first reported a
detection of PSR1706-44 at TeV energies in 1993 (\cite{kifune95})
based on sixty hours of observation in the summer of 1992. PSR1706-44
is identified with a pulsar (of period 102 ms) and appears to be
associated with a supernova remnant, possibly a plerion. At GeV
energies it has a very flat spectrum. The energy spectrum is hard
with a flux above 1 TeV of about $0.15 \times
10^{-11}$-cm$^{-2}$-s$^{-1}$. There is no evidence that the signal is
periodic. The detection has been confirmed by the University of Durham
group working in Narrabri, Australia (\cite{chadwick97}).

The CANGAROO group have also reported the detection of a 6$\sigma$
signal from the vicinity of the Vela pulsar (\cite{Yoshikoshi97}). The
integral $\gamma$-ray flux above 2.5 TeV is $2.5 \times 10^{-12}$
photons cm$^{-2}$ s$^{-1}$. Again there is no evidence for periodicity
and the flux limit is about a factor of ten less than the steady
flux. The signal is offset (by 0.14$^{\circ}$) from the pulsar
position which makes it more likely that the source is a synchrotron
nebula. Since this offset position is coincident with the birthplace
of the pulsar it is suggested that the progenitor electrons are relics
of the initial supernova explosion and they have survived because the
magnetic field was weak.

\subsection{Pulsars}
\label{pulsar-sec}

\begin{figure*}[t]
\centerline{\epsfig{file=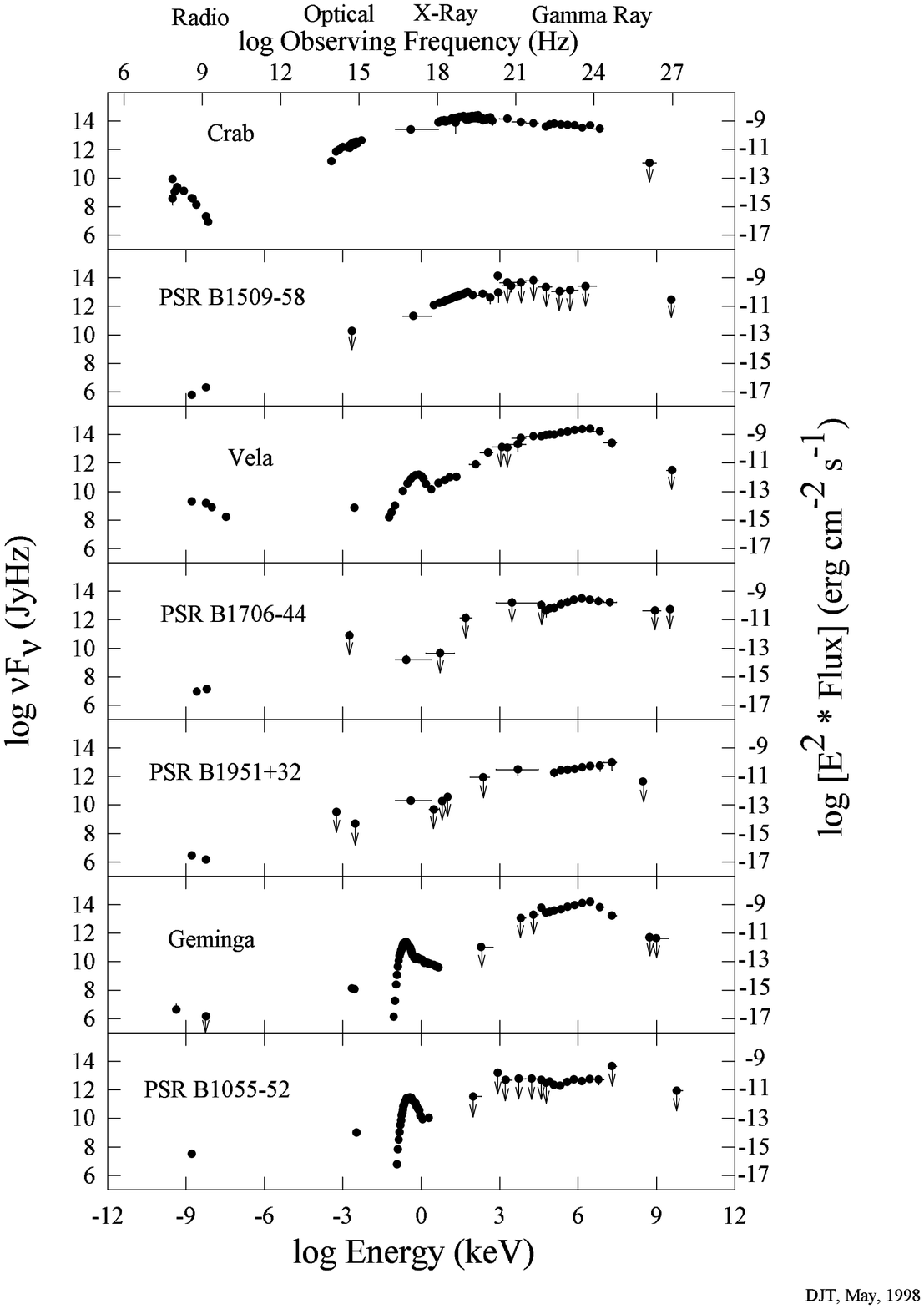,height=5.5in,angle=0.}}
\caption{Power spectrum of $\gamma$-ray pulsars detected with EGRET
(\protect\cite{thompson97}).
\label{pulsars-fig}}
\end{figure*}

The power spectra of most of the $\gamma$-ray pulsars
(\cite{thompson97}) are extremely flat with maximum power often coming
in the GeV energy range (see Figure~\ref{pulsars-fig}).  Because
pulsar models often involve electrons with energies up to 10$^{15}$eV,
it would come as no surprise if TeV $\gamma$-rays should emerge from
the pulsar magnetosphere and be detected. Although there is, as yet,
no established model for high energy $\gamma$-ray emission from
pulsars, it appears that, in general, the detection of pulsed TeV
$\gamma$-rays would favor outer gap (\cite{Romani96}) over polar cap
(\cite{Daugherty82}) models.  This is because, in the latter models,
the TeV $\gamma$-rays are attenuated by pair-production interactions
with the intense magnetic fields near the pulsars.

Sensitive upper limits have been obtained for emission by the Crab
pulsar (\cite{lessard99}), Geminga (\cite{akerlof93}) and the Vela
pulsar (\cite{Yoshikoshi97}); in general these confirm the steepening
of the spectra seen at 10\,GeV energies. The radio pulsar PSR\,1951+32
is particularly interesting because the power spectrum indicates that
the maximum power occurs at energies of at least a few GeV
(Figure~\ref{pulsars-fig}); in the EGRET measurements there is no
evidence for a high energy cut-off.  In fact, the flux continues to
rise with energy up to the highest energy observation.  Outer gap
models suggest that the pulsar should be detectable at higher
energies.  Observations of PSR\,1951+32 by the Whipple group reported
only an upper limit (\cite{srinivasan97}). This upper limit to the
pulsed flux is two orders of magnitude below the flux extrapolated
from the EGRET measurements. This represents the most dramatic
turn-over in the spectrum of a $\gamma$-ray pulsar and hence puts the
most severe constraints on the models.

\subsection{Supernova Remnants: Shell}
\label{snr}

Supernova remnants (SNRs) are widely believed to be the sources of
hadronic cosmic rays up to energies of approximately $Z \times
10^{14}$\,eV, where $Z$ is the nuclear charge of the particle (for a
review see \cite{Jones98}).  The arguments in support of this
statement are two-fold.  First, supernova blast shocks are some of the
few galactic sites capable of satisfying the energy required for the
production of galactic cosmic rays, although even these must have a
high efficiency, $\sim$10\% -- 30\% (e.g., Drury, Markiewicz \& V\"olk
1989), for converting the kinetic energy of supernova explosions into
high energy particles.  Second, the model of diffusive shock
acceleration (e.g., \cite{Blandford78}; \cite{Bell78};
\cite{Legage83}), which provides a plausible mechanism for efficiently
converting the explosion energy into accelerated particles, naturally
produces a power-law spectrum of ${\rm dN}/{\rm dE} \propto {\rm
E}^{-2.1}$.  This is consistent with the inferred spectral index at
the source for the observed local cosmic-ray spectrum of ${\rm
dN}/{\rm dE} \propto {\rm E}^{-2.7}$, after correcting for the effects
of propagation in the galaxy (e.g., \cite{Swordy90}).

The origins of cosmic rays cannot be studied directly because
interstellar magnetic fields isotropize their directions, except
perhaps at the highest energies ($\gtrsim 10^{18}$\,eV).  Thus, we
must look for indirect signals of their presence from astrophysical
sources.  That SNRs accelerate {\it electrons} to high energies is
well-established from observations of synchrotron emission in the
shells of SNRs at radio and, more recently, X-ray (e.g.,
\cite{Koyama95}) wavelengths.  However, it is difficult to extrapolate
from these data to inferences about the nature of the acceleration of
hadrons in these same objects.

Evidence of shock acceleration of hadronic cosmic rays in SNR shells
could come from measurements of $\gamma$-ray emission in these
objects.  Collisions of cosmic-ray nuclei with the interstellar medium
result in the production of neutral pions which subsequently decay
into $\gamma$-rays.  The $\gamma$-ray spectrum would extend from below
10\,MeV up to $\sim$1/10 of the maximum proton energy ($\gtrsim
10$\,TeV), with a distinctive break in the spectrum near 100\,MeV due
to the resonance in the cross-section for $\pi^0$ production.  As
$\gamma$-ray production requires interaction of the hadronic cosmic
rays with target nuclei, this emission should be stronger for those
SNR located near, or interacting with, dense targets, such as
molecular clouds.  The cosmic-ray density, and hence the associated
$\gamma$-ray luminosity, will increase with time as the SNR passes
through its free expansion phase, will peak when the SNR has swept up
as much interstellar material as contained in the supernova ejecta
(the Sedov phase) and gradually decline thereafter (Drury, Aharonian
\& V\"olk 1994; \cite{Naito94}).  Thus, $\gamma$-ray bright SNRs
should be ``middle-aged.''

From the calculations of Drury et al. (1994), the luminosity of
$\gamma$-rays from secondary pion production may be detectable by the
current generation of satellite-based and ground-based $\gamma$-ray
detectors, particularly if the objects are located in a region of
relatively high density in the interstellar medium.  In support of
this hypothesis, EGRET has detected signals from several regions of
the sky consistent with the positions of shell-type SNRs
(\cite{Sturner95}; \cite{Esposito96}; \cite{Lamb97}; \cite{Jaffe97}).
However, the EGRET detections alone are not sufficient to claim the
presence of high energy hadronic cosmic rays.  For instance, the
relatively poor angular resolution of EGRET makes difficult to
definitively identify the detected object with the SNR shell.  Because
of this, embedded pulsars (\cite{Brazier96}; \cite{deJager97}; Harrus,
Hughes \& Helfand 1996) and an X-ray binary (\cite{Kaaret99})
consistent with the positions of some of these EGRET sources have been
suggested as alternative counterparts.  In addition, significant
background from the diffuse Galactic $\gamma$-ray emission complicates
spectral measurements.  To complicate matters further, with the
detection of X-ray synchrotron radiation from SNR shells, the
possibility that $\gamma$-rays could be produced via inverse Compton
scattering of ambient soft photons has been realized
(\cite{Mastichiadis96b}; \cite{Mastichiadis96a}).  Bremsstrahlung
radiation may also be a significant source of $\gamma$-rays at MeV-GeV
energies (\cite{deJager97}; Gaisser, Protheroe \& Stanev 1998).

\subsubsection{VHE $\gamma$-ray observations}
\label{snr-obs-sec}

\begin{table*}[t]
\caption{Observations of shell-type supernova remnants \label{snruls}}
\begin{center}
\begin{tabular}{lrrrr} \hline\hline
 & \multicolumn{1}{c}{Observation} & & \multicolumn{1}{c}{Integral} & \\
\multicolumn{1}{c}{Object} & \multicolumn{1}{c}{Time} & 
 \multicolumn{1}{c}{Energy} & \multicolumn{1}{c}{Flux$^a$} & \\
\multicolumn{1}{c}{Name} & \multicolumn{1}{c}{(min.)} & 
 \multicolumn{1}{c}{(TeV)} & 
 \multicolumn{1}{c}{($10^{-11}$ cm$^{-2}$ s$^{-1}$)} & 
 \multicolumn{1}{c}{Ref.} \\ \hline
Tycho & 867.2 & $>$\,0.3 & $<$\,0.8 & \protect\cite{Buckley98b} \\
IC\,443$^b$ & 1076.7 & $>$\,0.3 & $<$\,2.1 & \protect\cite{Buckley98b} \\
            & 678.0 & $>$\,0.5 & $<$\,1.9$^c$ & \protect\cite{Hess97} \\
W\,44$^b$ & 360.1 & $>$\,0.3 & $<$\,3.0 & \protect\cite{Buckley98b} \\
W\,51 & 468.0 & $<$\,0.3 & $<$\,3.6 & \protect\cite{Buckley98b} \\
$\gamma$-Cygni$^b$ & 560.0 & $>$\,0.3 & $<$\,2.2 & \protect\cite{Buckley98b} \\
                  & 2820.0 & $>$\,0.5 & $<$\,1.1$^c$ & \protect\cite{Hess97} \\
W\,63 & 140.0 & $>$\,0.3 & $<$\,6.4 & \protect\cite{Buckley98b} \nl
SN\,1006 & 2040.0 & $>$1.7 & ${\rm 0.46 \pm 0.6_{stat} \pm 1.4_{sys}}$ &
 \protect\cite{Tanimori98a} \\ \hline
\multicolumn{5}{p{5.4in}}{$^a$Upper limits from Buckley et al. (1998) are at 
 the 99.9\% confidence level and those from Hess et al. (1997) are at the 
 3\,$\sigma$ confidence level.}\\
\multicolumn{5}{l}{$^b$Associated with an EGRET source.}\\
\multicolumn{5}{p{5.4in}}{$^c$Upper limits were converted from fractions of 
 $>$500\,GeV Crab flux using the measured Crab flux of Hillas et al. (1998).}\\
\end{tabular}
\end{center}
\end{table*}

Measurements of $\gamma$-rays at very high energies may help resolve
the puzzle of the $\gamma$-ray emission from the EGRET-detected
sources.  VHE $\gamma$-ray telescopes have much better angular
resolution than EGRET, reducing the source confusion associated with
any detection.  Also, because the diffuse Galactic $\gamma$-ray
emission has a relatively steep spectrum, $\propto {\rm E}^{-2.4} -
{\rm E}^{-2.7}$ (\cite{Hunter97}), compared with the expected
$\sim{\rm E}^{-2.1}$ spectrum of $\gamma$-rays from secondary pion
decay, contamination from background $\gamma$-ray emission should be
less in the VHE range.  Thus, in recent years, searches for emission
from shell-type SNRs have been a central part of the observation
program of VHE telescopes.

The Whipple Collaboration has published the results of observations of
six shell-type SNRs (IC\,443, $\gamma$-Cygni, W\,44, W\,51, W\,63, and
Tycho) selected as strong $\gamma$-ray candidates based on their radio
properties, distance, small angular size, and possible association
with a molecular cloud (\cite{Buckley98b};\cite{Hess97}).  The small
angular size was made a requirement due to the limited field of view
($3^\circ$ diameter) of the Whipple telescope at that time.  VHE
telescopes can also detect fainter $\gamma$-ray sources if they are
more compact, because they can reject more of the cosmic-ray
background.  IC\,443, $\gamma$-Cygni, and W\,44 are also associated
with EGRET sources (\cite{Esposito96}).  Despite long observations, no
significant excesses were observed, and stringent limits were derived
on the VHE flux (see Table~\ref{snruls}).

\begin{figure*}
\centerline{\epsfig{file=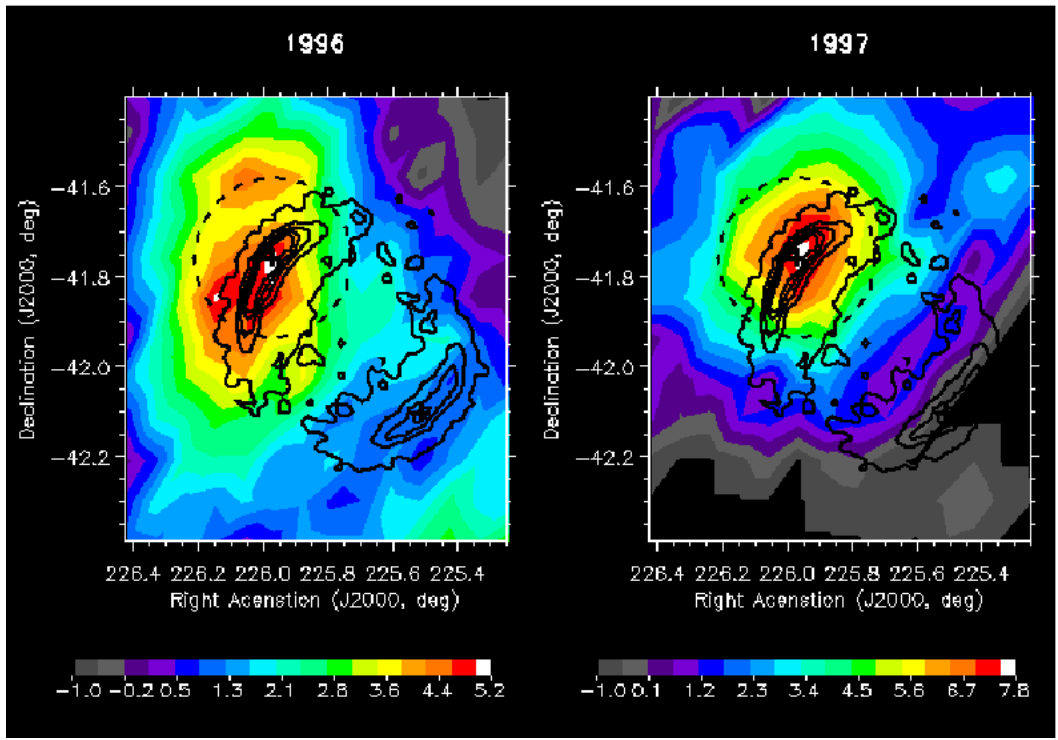,width=5.5in}}
\caption{Left: contour map of the statistical significance of the excess
emission from SN\,1006 as observed with the CANGAROO telescope in
1996.  Right: the same plot for data taken in 1997.  The solid lines
indicate the contour map of non-thermal X-ray emission detected with
{\it ASCA}.  The dashed circles indicate the angular resolution of the
CANGAROO telescope.  Figure from Tanimori et al. (1998a).
\label{sn1006_tev}
}
\end{figure*}

In contrast to the upper limits derived by the northern hemisphere
telescopes, the CANGAROO Collaboration has recently reported evidence
for TeV $\gamma$-ray emission from the shell-type SNR, SN\,1006
(\cite{Tanimori98a}).  Observations taken in 1996 and 1997 indicate a
statistically significant excess from the northeast rim of the SNR
shell (see Figure~\ref{sn1006_tev}).  The position of the excess is
consistent with the location of non-thermal X rays detected by the
{\it ASCA} experiment (\cite{Koyama95}).  If this object is confirmed
as a TeV $\gamma$-ray source, it represents the first direct evidence
of acceleration of particles to TeV energies in the shocks of SNRs.

\placefigure{sn1006_tev}

\subsubsection{Implications of the $\gamma$-ray observations}
\label{snrimps}

\begin{figure*}[t]
\centerline{\epsfig{file=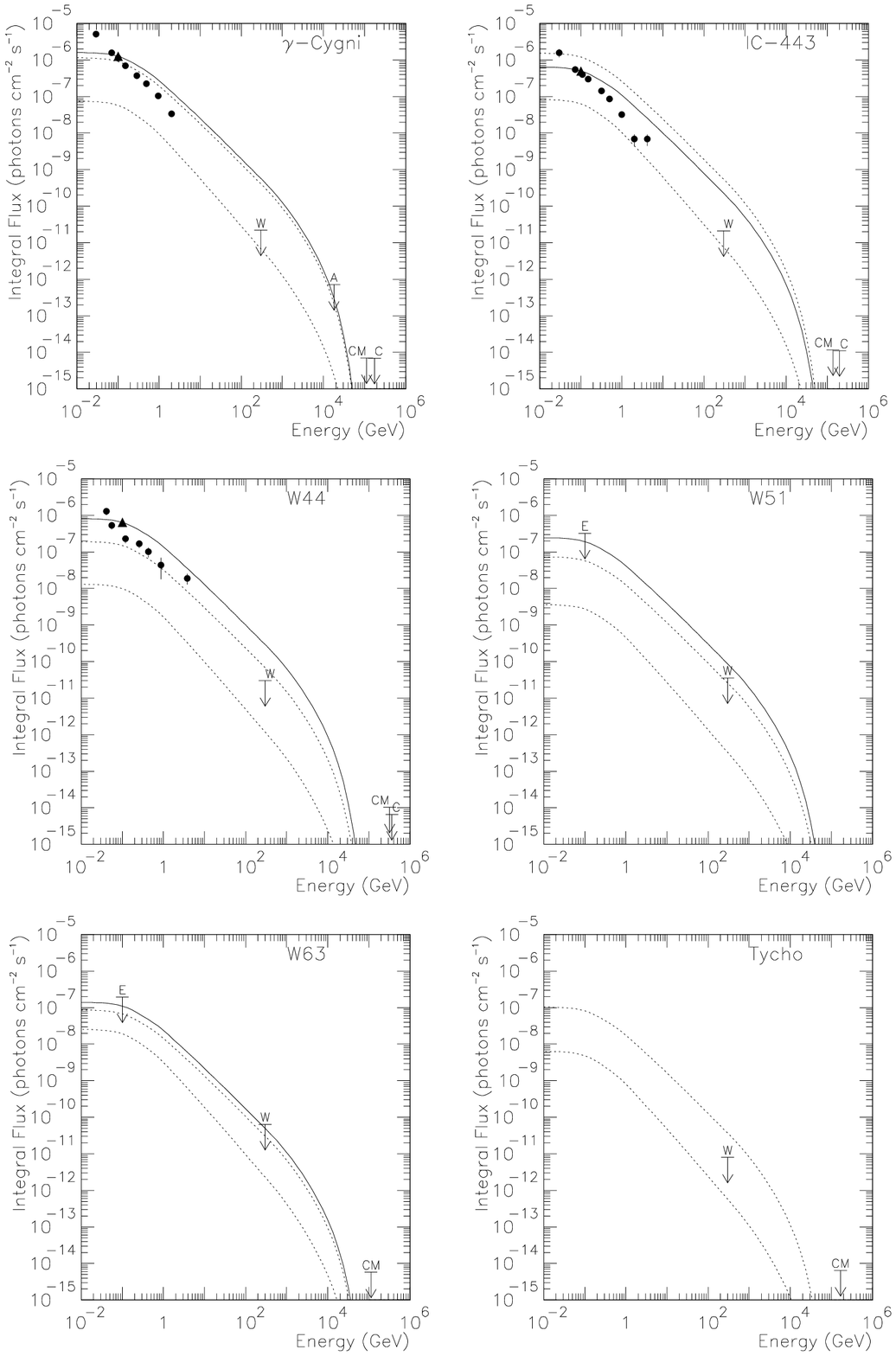,height=6in}}
\caption{Whipple Observatory upper limits (W) shown along with EGRET
integral fluxes (E) and integral spectra.  Also shown are CASA-MIA
upper limits (CM) from Borione et al. (1995), Cygnus upper limits (C)
from Allen et al. (1995), and the AIROBICC upper limit from Prosch et
al. (1996).  The solid curves indicate extrapolations from the EGRET
integral data points at 100\,MeV (indicated by the triangles).  The
dashed curves are estimates of the allowable range of fluxes from the
model of Drury et al. (1994).  Figure from Buckley et al. (1998).
\label{snrlims_fig}
}
\end{figure*}

If the EGRET detections do indicate the presence of $\gamma$-rays
produced by secondary pion decay, the measured flux can be compared
with the VHE upper limits.  These VHE upper limits and the EGRET
measurements are compared to the predicted fluxes from the model of
Drury et al. (1994) in Figure~\ref{snrlims_fig} (\cite{Buckley98b}).
The solid curves are normalized to the integral $>$100\,MeV flux
detected by EGRET assuming a source cosmic-ray spectrum of ${\rm
E}^{-2.1}$, so they assume that the EGRET emission is entirely due to
cosmic-ray interactions and that the cosmic-ray spectrum at the source
has the canonical spectrum.  In the cases of $\gamma$-Cygni, IC\,443,
and W\,44, the Whipple upper limits lie a factor of $\sim$25, 10, and
10, respectively, below the model extrapolations and require either a
spectral break or a differential source spectrum steeper than ${\rm
E}^{-2.5}$ for $\gamma$-Cygni and ${\rm E}^{-2.4}$ for IC\,443
(\cite{Buckley98}).  In addition, the measured EGRET spectra, while
consistent with a spectral index of about 2, do not show the
flattening of the $\gamma$-ray spectrum near 100\,MeV which would be
expected if the $\gamma$-rays result from secondary pion decay.
Gaisser et al. (1998) performed multiwavelength fits to the EGRET and
Whipple results and concluded that if the EGRET detections are truly
from the shells of the SNRs, the EGRET data must be dominated at low
energies by electron bremsstrahlung radiation, but the source spectrum
at high energies must still be relatively steep ($\sim$E$^{-2.4}$) to
account for the Whipple upper limits (cf., \cite{Buckley98b}).

If, on the other hand, the $>$100\,MeV emission results from some
other emission mechanism, such as the interactions of high energy
electrons accelerated by embedded pulsars or in the SNR shocks, the
EGRET results should not be compared to the Whipple data.  Instead,
the Whipple data must be considered alone in the context of the
secondary pion decay models.  This possibility was also investigated
by Buckley et al. (1998) and is shown by the dashed curves in
Figure~\ref{snrlims_fig}.  The dashed curves represent a conservative
estimate of the range of allowable parameter values for the Drury et
al. (1994) model, without reference to the EGRET detections.  For this
comparison, there is still room for the models to work in these
objects.  However, the upper limits in some (e.g., IC\,443) are
beginning to strain the limits of the available parameter space.  It
will take more sensitive measurements with future telescopes to fully
span the allowable parameter space and see if we need to reconsider
our assumptions about the sources of cosmic rays within our Galaxy.

The TeV emission from SN\,1006 detected by the CANGAROO group also
does not require the presence of hadronic cosmic rays.  In fact, the
most common explanation for the detected emission is inverse Compton
scattering of electrons with cosmic microwave background photons
(\cite{Reynolds96}; \cite{Mastichiadis96b}).  The main arguments for
this are that the emission is centered on one of the regions where the
synchrotron emission was detected with {\it ASCA} and the lack of
evidence for a nearby molecular cloud needed to boost the TeV
$\gamma$-ray flux to a detectable level.  Under the assumption that
the emission is from inverse Compton scattering, Tanimori et
al. (1998a) have combined their data with the {\it ASCA} results to
derive an estimate of 6.5 $\pm$ 2\,$\mu$G for the magnetic field
within the SNR shell.  The observations also provide an upper limit on
the acceleration time.  Thus, the TeV observations provide previously
unknown parameters for models of the shock acceleration in SNRs.
Unfortunately, the possibility of inverse Compton emission from the
shells of SNRs also confuses the issue for using $\gamma$-rays as a
probe of cosmic-ray acceleration in SNRs.  Future measurements will
need to provide accurate spectra and spatial mapping of the
$\gamma$-ray emission from SNRs in order for the source of the
$\gamma$-ray emission to be unambiguously resolved.

\section{Extragalactic Sources}
\label{extrag-sec}

\subsection{Active Galactic Nuclei}
\label{AGN-sec}

In recent years, high energy $\gamma$-rays have come to play an
important role in the study of AGNs.  Before the launch of {\it CGRO}
in 1991, the only known extragalactic source of high energy
$\gamma$-rays was 3C\,273 which had been detected with the COS-B
satellite 20 years ago (\cite{Swanenburg78}).  The EGRET detector on
the {\it CGRO} has identified more than 65 AGNs which emit
$\gamma$-rays at energies above 100\,MeV (\cite{Hartman99}), and a
substantial fraction of those sources which remain unidentified in the
EGRET catalog are likely to be AGNs as well.  In addition, the Whipple
Observatory $\gamma$-ray telescope has discovered three AGNs which
emit at energies above 300\,GeV (\cite{Punch92}; \cite{Quinn96};
\cite{Catanese98}) and there are recent detections of two other AGNs
with Cherenkov telescopes (\cite{Chadwick99}; \cite{Neshpor98}).
During flaring episodes, the $\gamma$-ray emission can greatly exceed
the energy output of the AGNs at all other wavelengths.  Thus, any
attempt to understand the physics of these objects must include
consideration of the $\gamma$-ray emission.

All of the AGNs detected in high energy $\gamma$-rays are radio-loud
sources with the radio emission arising primarily from a core region
rather than from lobes.  These types of AGNs are often collectively
referred to as ``blazars,'' and include BL Lacertae (BL Lac) objects,
flat spectrum radio-loud quasars (FSRQs), optically violent variables,
and superluminal sources.  The emission characteristics of blazars
include high polarization at radio and optical wavelengths, rapid
variability at all wavelengths, and predominantly non-thermal emission
at most wavelengths.  The emission from blazars is believed to arise
from relativistic jets oriented at small angles to our line of sight.
If so, the observed radiation will be strongly amplified by
relativistic beaming (\cite{Blandford78b}).  Direct evidence for
relativistic beaming of the radio emission comes from Very Long
Baseline Interferometer (VLBI) observations of apparent superluminal
motion in many blazars (e.g., \cite{Vermeulen94}).  The rapid
variability and high luminosities of the detected $\gamma$-ray sources
imply that the $\gamma$-rays are also beamed (see \S~\ref{agnimps}).

There is a growing consensus that blazars are all the same type of
object, perhaps differing only in intrinsic luminosity (e.g.,
\cite{Fossati98}; \cite{Ghisellini98}) or some combination of
luminosity and viewing angle (e.g., \cite{Georganopoulos98}).
However, for this work we will continue the practice of referring to
BL Lac objects and FSRQs as distinct objects: BL Lac objects are those
blazars which have optical emission lines with equivalent width
$<5$\,\AA\ and FSRQs are the remaining blazars.  As we will see, this
distinction may be important in explaining why only BL Lac objects are
detected at very high energies.

\begin{figure*}[t]
\begin{minipage}[t]{3.4in}
\centerline{\epsfig{file=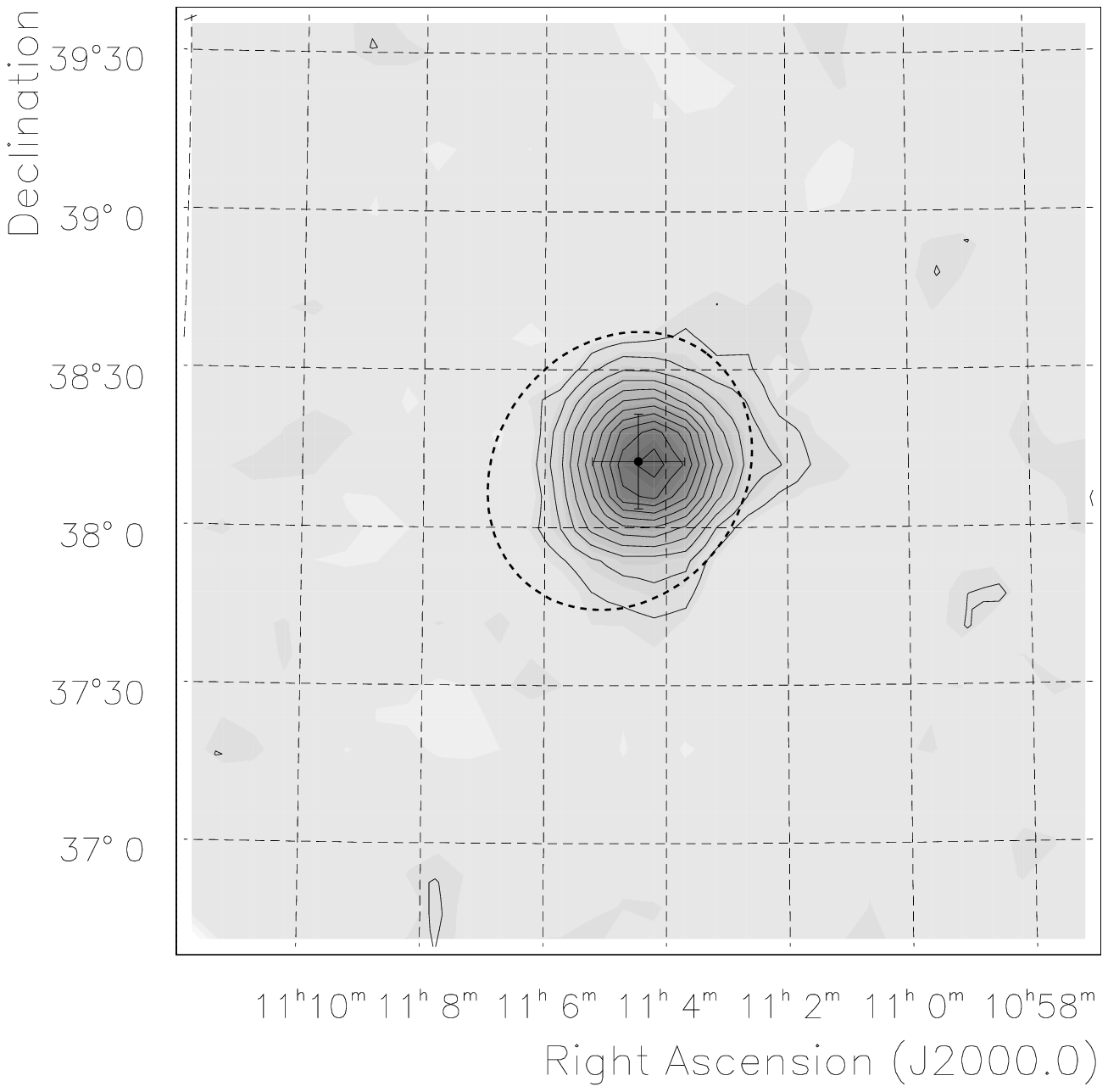,width=3.3in}}
\caption{Two-dimensional plot of the VHE $\gamma$-ray emission from
the region around Mrk\,421.  The gray scale is proportional to the
number of excess $\gamma$-rays and the solid contours correspond to
2\,$\sigma$ levels.  The dashed ellipse give the 95\% confidence
interval determined by EGRET (\protect\cite{Thompson95}).  The
position of Mrk\,421 is indicated by the cross.  Figure from Buckley
et al. (1996).
\label{m4-2d-fig}
}
\end{minipage}
\hfill
\begin{minipage}[t]{3.4in}
\centerline{\epsfig{file=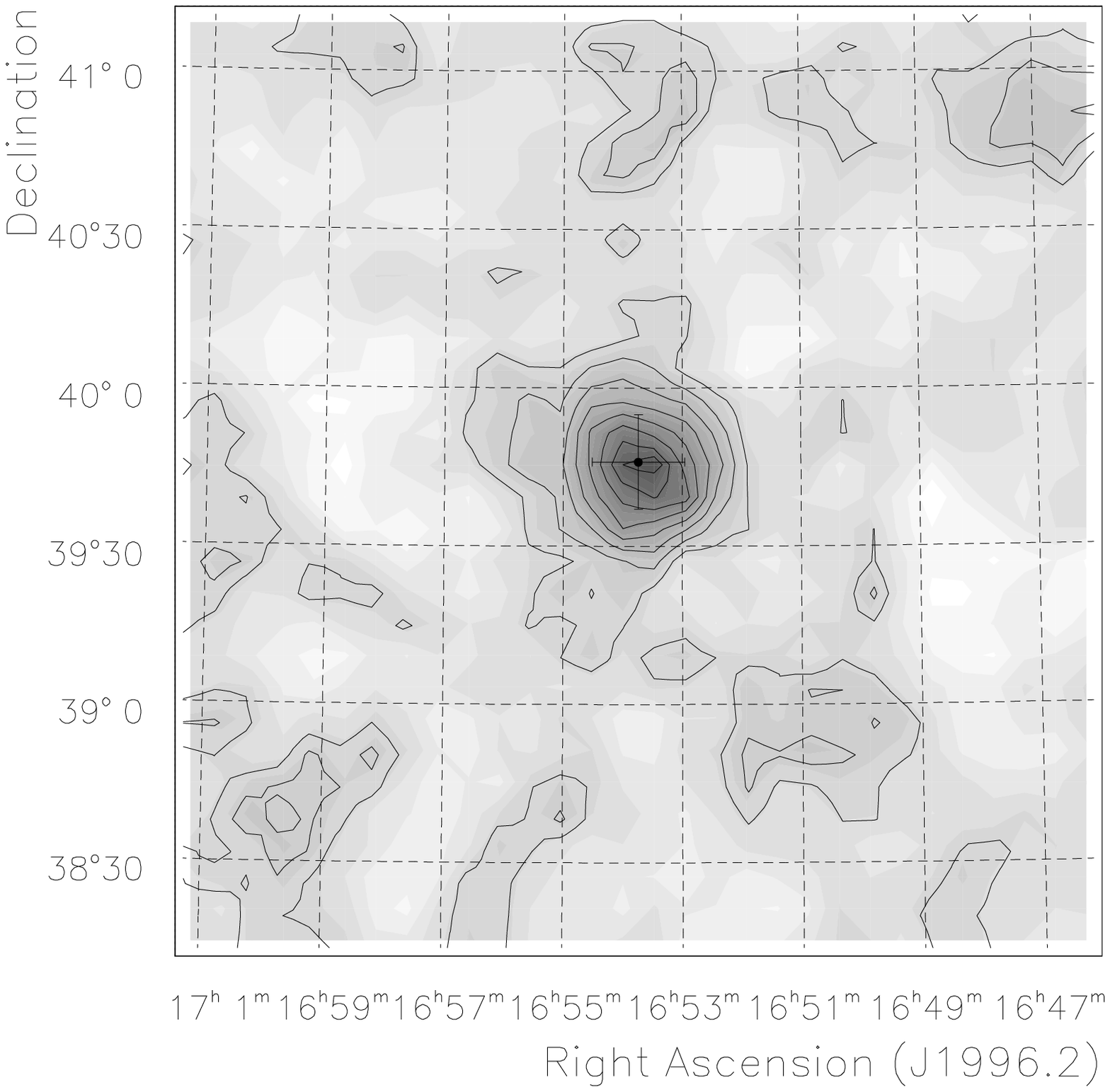,width=3.3in}}
\caption{Two-dimensional plot of the VHE $\gamma$-ray emission from
the region around Mrk\,501.  The gray scale is proportional to the
number of excess $\gamma$-rays and the solid contours correspond to
1\,$\sigma$ levels.  The position of Mrk\,501 is indicated by the
cross.
\label{m5-2d-fig}
}
\end{minipage}
\end{figure*}

The spectral energy distribution of blazars appears to consist of two
parts.  First, a low energy component exhibits a power per decade
distribution that rises smoothly from radio wavelengths up to a broad
peak in the range spanning infrared (IR) to X-ray wavelengths,
depending on the specific blazar type, above which the power output
rapidly drops off.  Second, a distinct, high energy component, which
does not extend smoothly from the low energy component, is often seen.
It typically becomes apparent in the X-ray range and has a peak power
output in the $\gamma$-ray range between $\sim$1\,MeV and 1\,TeV
(e.g., \cite{Montigny95}), again depending on the specific blazar
type.  When plotted as E$^2$dN/dE (or equivalently $\nu$F$_\nu$) the
spectral energy distribution shows a two-humped shape, though some
objects show evidence of a third component (e.g., \cite{Kubo98}).

Though there is no general consensus on the origin of these emission
components, it is generally agreed that the low energy component
arises from incoherent synchrotron emission by relativistic electrons
within the jet (e.g., \cite{Blandford78b}).  This is supported most
strongly by the high-level, variable polarization observed in these
objects at radio and optical wavelengths.  The origin of the high
energy emission is a matter of great interest.  There are many
variations of the models and here we only briefly mention a few which
are most often invoked to explain the $\gamma$-ray emission.  The most
popular models at this time are those in which the $\gamma$-rays are
produced through inverse Compton scattering of low energy photons by
the same electrons which produce the synchrotron emission at lower
energies.  Synchrotron self-Compton (SSC) emission (e.g.,
\cite{Konigl81}; \cite{Maraschi92}; \cite{Bloom96}), in which the seed
photons for the scattering are the synchrotron photons already present
in the jet, must occur at some level in all blazars, but models in
which the $\gamma$-ray emission arises predominantly from inverse
Compton scattering of seed photons which arise outside of the jet,
either directly from an accretion disk (Dermer, Schlickeiser \&
Mastichiadis 1992) or after being re-processed in the broad-line
region or scattering off thermal plasma (Sikora, Begelman \& Rees
1994), appear to fit the observations satisfactorily as well.  Another
set of models proposes that the $\gamma$-rays are produced by
proton-initiated cascades (e.g., \cite{Mannheim93}).  As we will show
in \S\ref{agnimps}, the $\gamma$-ray observations strain both types of
models, but do not, at present, rule any out.

In the remainder of this section we discuss the status of VHE
observations and the emission characteristics of the detected objects
(\S\ref{vhestat}), the results of multi-wavelength campaigns on the
detected objects (\S\ref{multi}) which are the best probe of the
physics of blazars, and finally briefly discuss some of the
implications of these observations on our understanding of the physics
of the blazars and on the models which purport to explain them.

\subsubsection{Observational Status and Emission Characteristics}
\label{vhestat}

The BL Lac object Markarian\,421 (Mrk\,421, $z=0.031$) was detected as
the first extragalactic source of VHE $\gamma$-rays in 1992 using the
Whipple Observatory $\gamma$-ray telescope (\cite{Punch92}).  A
two-dimensional image of the emission from Mrk\,421 is shown in
Figure~\ref{m4-2d-fig}.  Although Mrk\,421 had previously been part of
an active program of observing extragalactic sources (\cite{Cawley85})
by the Whipple Collaboration, the observations which led to the
detection of Mrk\,421 at TeV energies were initiated in response to
the detection of several AGN by the EGRET experiment.  The initial
detection indicated a 6\,$\sigma$ excess and the flux above 500\,GeV
was approximately 30\% of the flux of the Crab Nebula at those
energies.  Mrk\,421 has been confirmed as a source of VHE
$\gamma$-rays by the HEGRA Collaboration (\cite{Petry96}), the
Telescope Array Project (\cite{Aiso97}), and the SHALON telescope
(\cite{Sinitsyna97}) and, as discussed in \S\,\ref{multi} below,
multi-wavelength correlations have confirmed that the VHE source is
indeed Mrk\,421 and not some other object.

\begin{figure*}[t]
\centerline{\epsfig{file=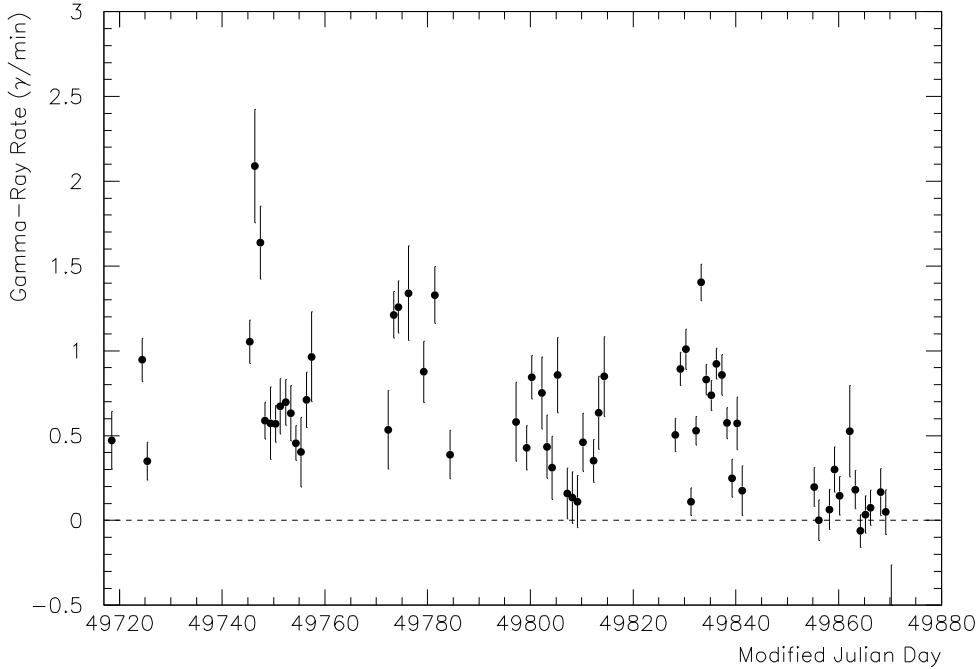,height=3.5in}}
\caption{Daily VHE $\gamma$-ray count rates for Mrk\,421 during
1995.
Modified Julian Day 49720 corresponds to 1995 January 3.  Figure from
Buckley et al. (1996).
\label{m4_1995_lc}
}
\end{figure*}

With the successful detection of Mrk\,421, the Whipple Collaboration
initiated a search for VHE emission from several other blazar-type
AGNs, concentrating at first on those objects detected by EGRET, but
also spending a substantial amount of time observing radio-loud
blazars which were not detected by EGRET.  This broad approach led to
the detection of the BL Lac object Mrk\,501 ($z=0.034$) by the Whipple
Collaboration in 1995 (\cite{Quinn96}).  A two-dimensional map of the
VHE emission from Mrk\,501 is shown in Figure~\ref{m5-2d-fig}.
Because Mrk\,501 had not been detected as a significant source of
$\gamma$-rays by EGRET, this was the first object to be discovered as
a $\gamma$-ray source from the ground.  Hence, VHE $\gamma$-ray
astronomy was established as a legitimate channel of astronomical
investigations in its own right, not just an adjunct of high energy
observations from space.  The flux of Mrk\,501 during 1995 was, on
average, 10\% of the VHE flux of the Crab Nebula, making it the
weakest detected source of VHE $\gamma$-rays.  Mrk\,501 was confirmed
as a source of VHE $\gamma$-rays in 1996 by the HEGRA  telescopes
(\cite{Bradbury97}),  the CAT telescope
(\cite{Punch97}), the Telescope Array Project (\cite{Hayashida98}),
and TACTIC (\cite{Bhat97}).

In addition to the confirmed detections of Mrk\,421 and Mrk\,501,
three other objects have recently been reported as sources of VHE
$\gamma$-rays, but remain to be verified by detections from independent
$\gamma$-ray telescopes.  The BL Lac object 1ES\,2344+514 ($z=0.044$)
was detected at energies about 350\,GeV by the Whipple Observatory in
1995 (\cite{Catanese98}).  Most of the emission comes from a single
night, December 20, in which a flux of approximately half that of the
Crab Nebula was detected with a significance of 6\,$\sigma$.  Other
observations during that year revealed an excess of
4\,$\sigma$; subsequent observations have yielded no significant
signal.  1ES\,2344+514 is not detected by EGRET (\cite{Thompson96a})
so if this detection is confirmed it is another instance of a
$\gamma$-ray source being  first detected by a ground-based telescope.
PKS\,2155-304 ($z=0.117$), often considered the archetypical X-ray
selected BL Lac object, was detected at energies above 300\,GeV at the
7\,$\sigma$ level by combining observations from 1996 and 1997 by the
Durham group (\cite{Chadwick99}).  The flux was
approximately 40\% of the VHE flux of the Crab Nebula and
corresponded to an active X-ray emission period.  PKS\,2155-304 is an
EGRET source with an average flux at $E > 100$\,MeV comparable to that
of Mrk\,421.  Finally, the BL Lac object 3C\,66A ($z=0.444$) has been
reported as a source of $>$900\,GeV $\gamma$-rays based on a
5\,$\sigma$ excess seen in observations in 1996 by $\gamma$-ray
telescopes at the Crimean Astrophysical Observatory
(\cite{Neshpor98}).  The average flux during these observations was
approximately 120\% of the flux of the Crab Nebula at these energies.
3C\,66A is an EGRET source (\cite{Hartman99}).

Extreme variability on time-scales from minutes to years is the most
distinctive feature of the VHE emission from these BL Lac objects.
Variability in the emission is a surprising feature in some respects
because it implies a small emission region.  If low energy photons
(e.g., infrared, optical, and ultraviolet) are produced in the same
region, the VHE photons would pair produce with these photons and
would not escape.  Also, if the variability occurs near the base of
the jet, there is likely to be considerable ambient radiation
present which can attenuate the $\gamma$-ray signal.  This opacity
problem is reduced considerably if the emission is beamed toward us
(e.g., \cite{Dermer95}; \cite{Buckley96}), and this has been one of
the main arguments for $\gamma$-ray beaming in these objects
(see \S\,\ref{agnimps}).

\begin{figure*}[t]
\centerline{\epsfig{file=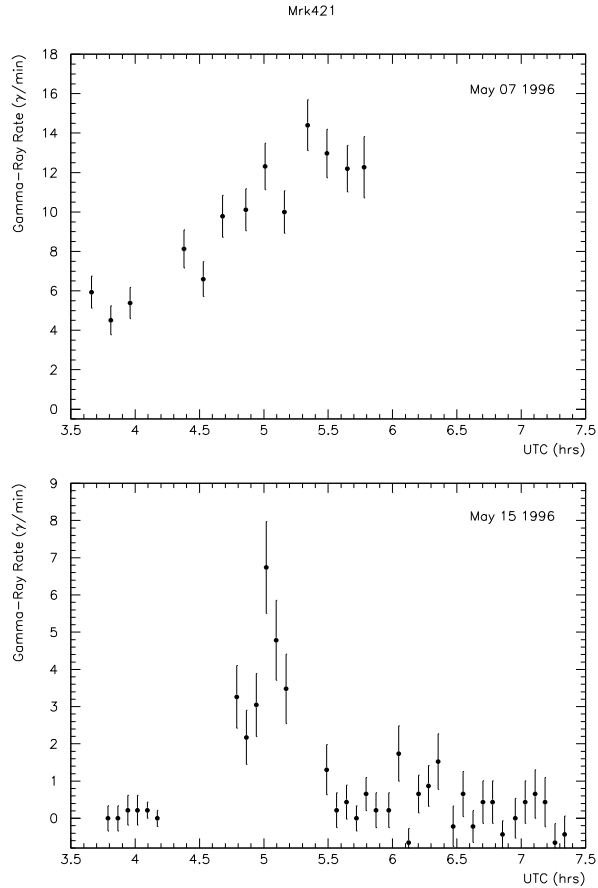,height=5in}}
\caption{Lightcurves of two flares observed from Mrk\,421 by the
Whipple Collaboration on 1996 May 7 (a) and May 15 (b).  The time axes
are shown in coordinated universal time (UTC) in hours.  For the May 7
flare, each point is a 9-minute integration; for the May 15 flare, the
integration time is 4.5 minutes.  Figure from Gaidos et al. (1996).
\label{m4_fastvars}
}
\end{figure*}

The first clear detection of flaring activity in the VHE emission of
an AGN came in 1994 observations of Mrk\,421 by the Whipple
Collaboration (\cite{Kerrick95a}) where a 10-fold increase in the
flux, from an average level that year of approximately 15\% of the
Crab flux to approximately 150\% of the Crab flux, was observed. 
Subsequent analysis of Mrk 421 data indicated evidence for
less prominent episodes of variability during 1992 and 1993 as well
(\cite{Schubnell96}).  This suggested that variability could be
present on a fairly frequent basis in the VHE emission.  In order to
characterize this variability, the Whipple Collaboration began
systematic monitoring of Mrk\,421 in 1995 which continues to the
present day.  The observations of Mrk\,421 in 1995, shown in
Figure~\ref{m4_1995_lc}, revealed several distinct episodes of flaring
activity, like in previous observations, but, perhaps more
importantly, indicated that the VHE emission from Mrk\,421 was best
characterized by a succession of day-scale or shorter flares with a
baseline emission level below the sensitivity limit of the Whipple
detector (\cite{Buckley96}).  The time-scale of the flaring is derived
from the fact that, for the most part, the flux levels measured each
night varied fairly randomly with no evidence of a smooth pattern.
Thus, though no significant intra-night variability was discerned in
these observations, it seemed clear that it could occur.

The hypothesis that the VHE emission of Mrk\,421 could flare on
sub-day time-scales was borne out in spectacular fashion in 1996, with
the observations of two flares by the Whipple Collaboration
(Figure~\ref{m4_fastvars}; \cite{Gaidos96}).  In the first flare,
observed on May 7, the flux increased monotonically during the course
of $\sim$2 hours of observations, beginning at a rate twice as high as
any previously observed flare and reaching a counting rate $\approx$10
times the rate from the Crab, at which point observations had to stop
because of moonrise.  This flux is the highest observed from any VHE
source to date.  The doubling time of the flare was $\sim$1 hour.
Follow-up observations on May 8 showed that the flux had dropped to a
flux level of $\approx$30\% of the Crab Nebula flux, implying a
decay time-scale of $<$1 day.  The second flare, observed on May 15,
although weaker was remarkable for its very short duration: the entire
flare lasted approximately 30 minutes with a doubling and decay time
of less than 15 minutes.  {\it These two flares are the fastest time-scale
variability, by far, seen from any blazar at any $\gamma$-ray energy.}

\begin{figure*}[t]
\centerline{\epsfig{file=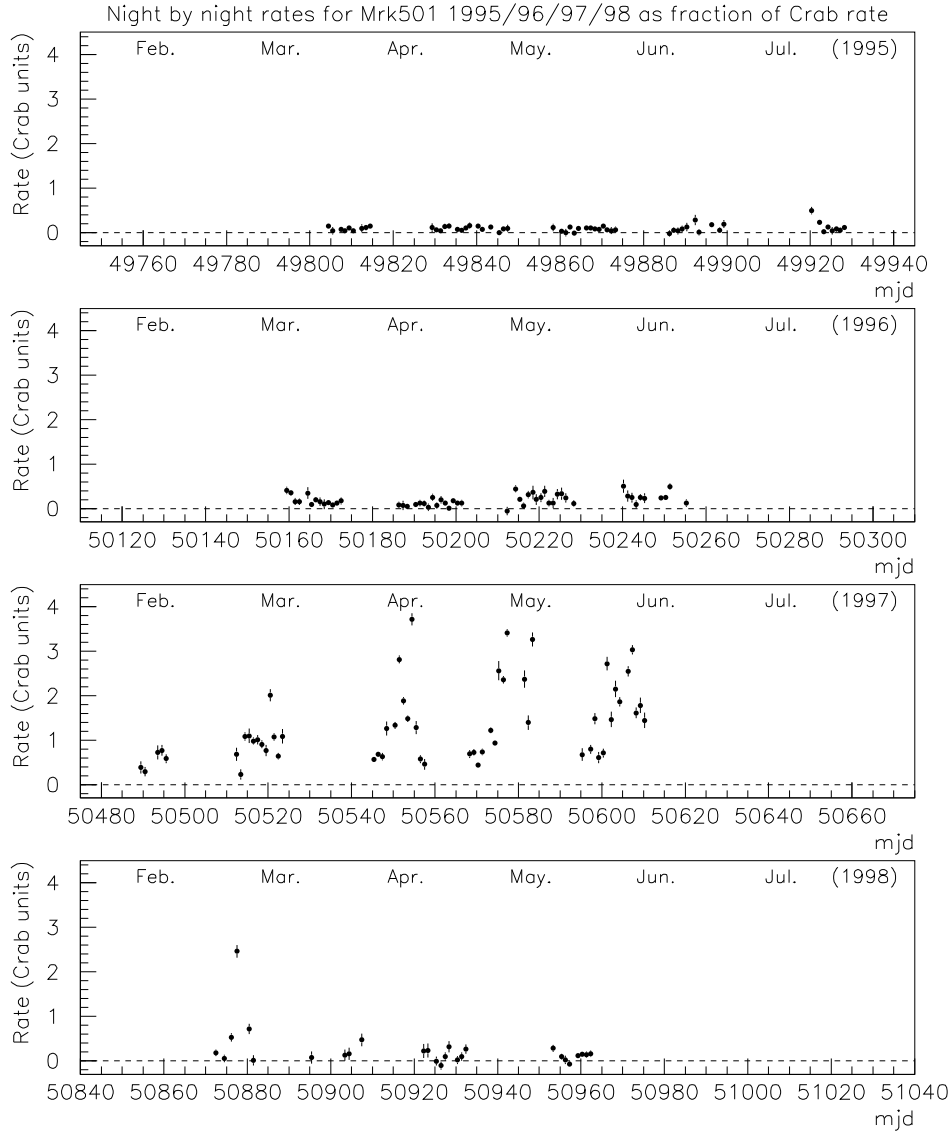,height=6in}}
\caption{VHE $\gamma$-ray lightcurve for Mrk\,501 as observed with the
Whipple telescope between 1995 and 1998 at energies above 350\,GeV.
Fluxes are expressed as fractions of the Crab Nebula flux above
350\,GeV.  Figure adapted from Quinn et al. (1999).
\label{m5-4yr-fig}
}
\end{figure*}

Systematic observations of Mrk\,501 sensitive to day-scale flares have
been conducted since 1995 with the Whipple Observatory $\gamma$-ray
telescope (\cite{Quinn99}) and since 1997 with the telescopes of the
HEGRA (\cite{Aharonian99a}), CAT (\cite{Punch97}), and Telescope Array
(\cite{Hayashida98}) collaborations.  The results of these
observations indicate a wide range of emission levels
(Figure~\ref{m5-4yr-fig}) and some very interesting similarities and
differences with the VHE emission from Mrk\,421.  The observations in
1995 indicate a flux which is constant, with the exception of one
night, MJD 49920, when the flux was approximately 4.6\,$\sigma$ above
the average (approximately 5 times the flux during the remainder of
the season) (\cite{Quinn96}; \cite{Quinn99}).  Observations in 1996 by
the Whipple Observatory show that the average flux of Mrk\,501 had
increased to approximately 20\% of the Crab Nebula flux above 300\,GeV
indicating a two-fold increase in the average flux over the 1995
observations (\cite{Quinn99}).  HEGRA observations indicated an
average flux of approximately 30\% of the Crab flux above 1.5\,TeV,
perhaps indicating a harder emission spectrum than that of the Crab
Nebula.  The Whipple observations show no clear flaring episodes but
the probability that the average monthly flux levels are drawn from a
distribution with a constant flux level is 3.6$\times$10$^{-5}$,
clearly indicating that the emission is varying on at least
month-scales (\cite{Quinn99}).  There is no significant evidence for
day-scale variations within each month in 1996.

\begin{figure*}[t]
\centerline{\epsfig{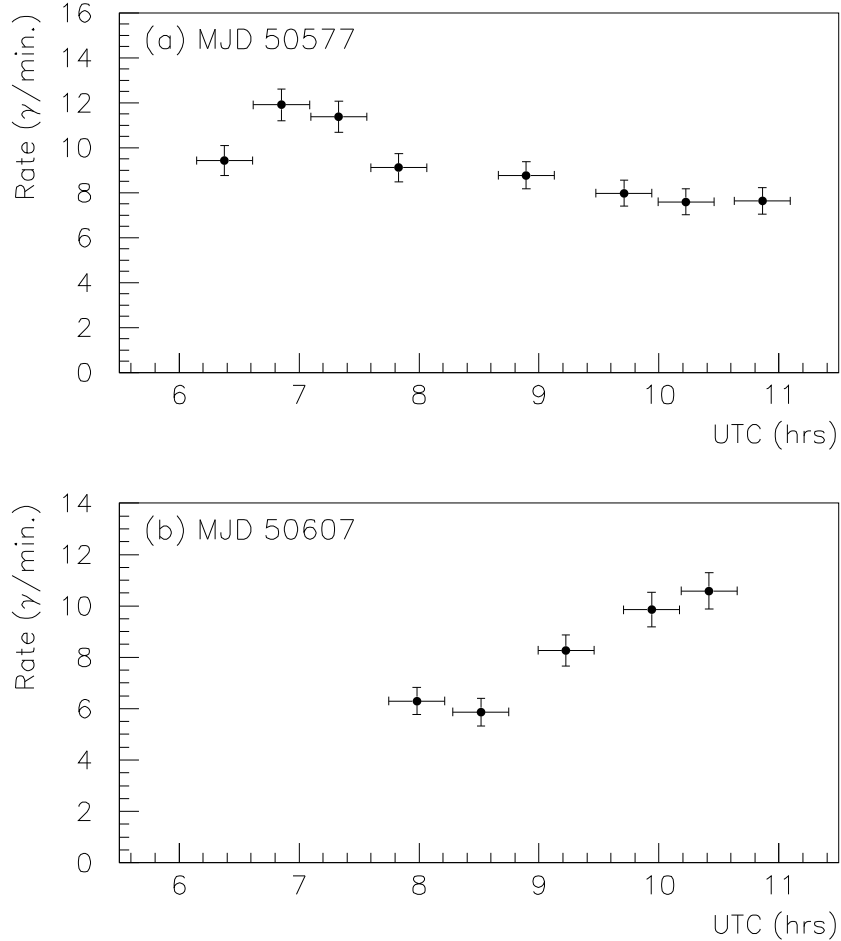}}
\caption{Very high energy $\gamma$-ray light curves of Mrk\,501 for
the two nights in 1997 which show significant intra-night variability.
Figure from Quinn et al. (1999).
\label{m5_hourly}
}
\end{figure*}

In 1997, the VHE emission from Mrk\,501 changed dramatically.  After
being the weakest known source in the VHE sky in 1995 and 1996, it
became the brightest, with an average flux greater than that of the
Crab Nebula (whereas previous observations had never revealed a flux
$>$50\% of the Crab flux).  Also, the amount of day-scale flaring
increased and, for the first time, significant hour-scale variations
were seen.  Two clear episodes of hour-scale variability were detected
with the Whipple Observatory telescope (Figure~\ref{m5_hourly}) and a
search for intraday variability revealed several other nights which,
considered alone, would not have been considered significant but, when
combined, indicated frequent intra-day variability which was just
below the sensitivity of the Whipple telescope (\cite{Quinn99}).
Analysis of data from the HEGRA (\cite{Aharonian99a}) and Telescope
Array (\cite{Hayashida98}) projects revealed no statistically
significant intra-night variations, but the two nights in the HEGRA
data with the smallest statistical probability of having constant
emission are the same nights seen to have significant variability in
the Whipple data.

\begin{figure*}[t]
\centerline{\epsfig{file=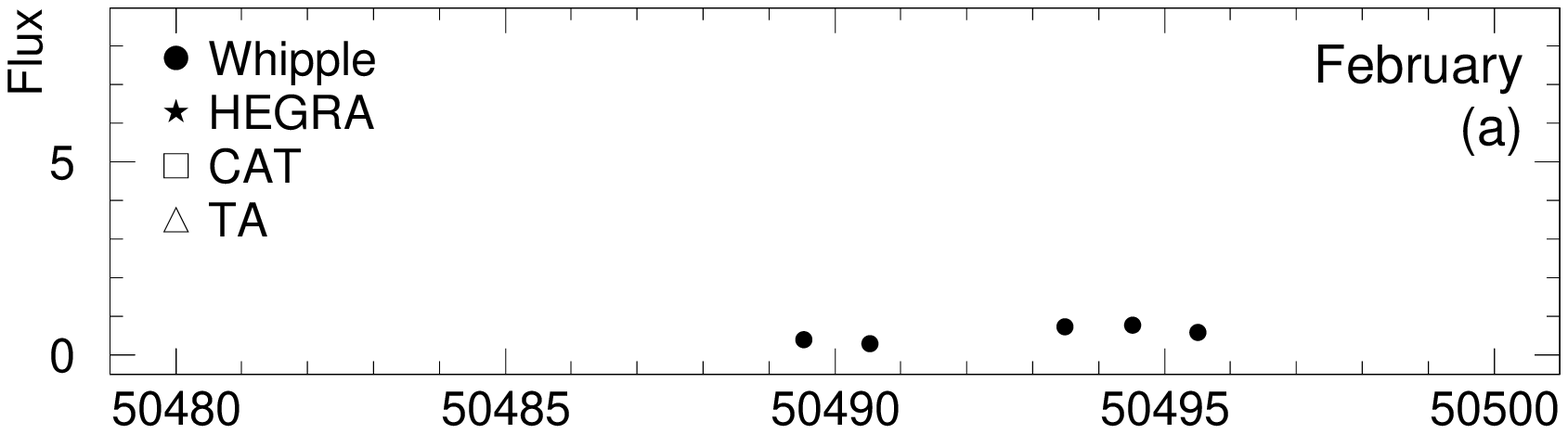,width=3.1in}\hspace*{0.1in}\epsfig{file=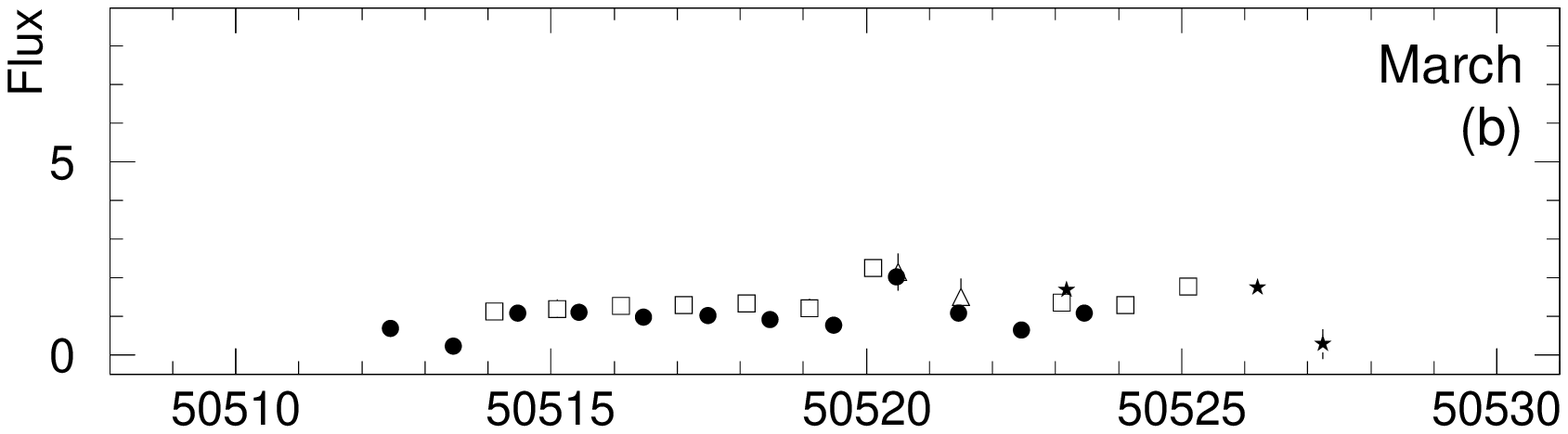,width=3.1in}}
\centerline{\epsfig{file=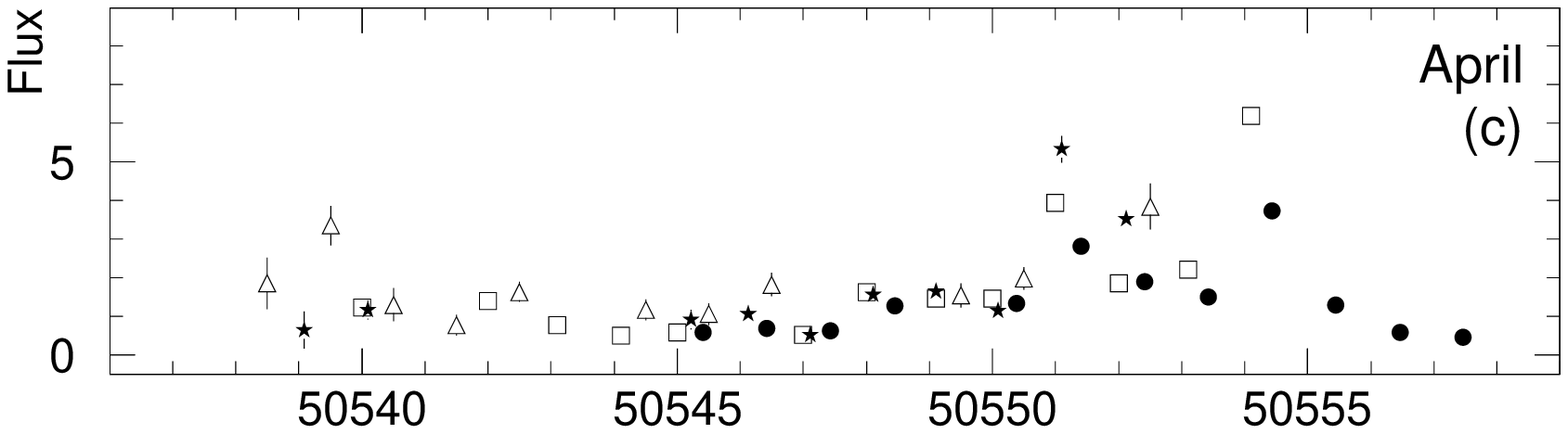,width=3.1in}\hspace*{0.1in}\epsfig{file=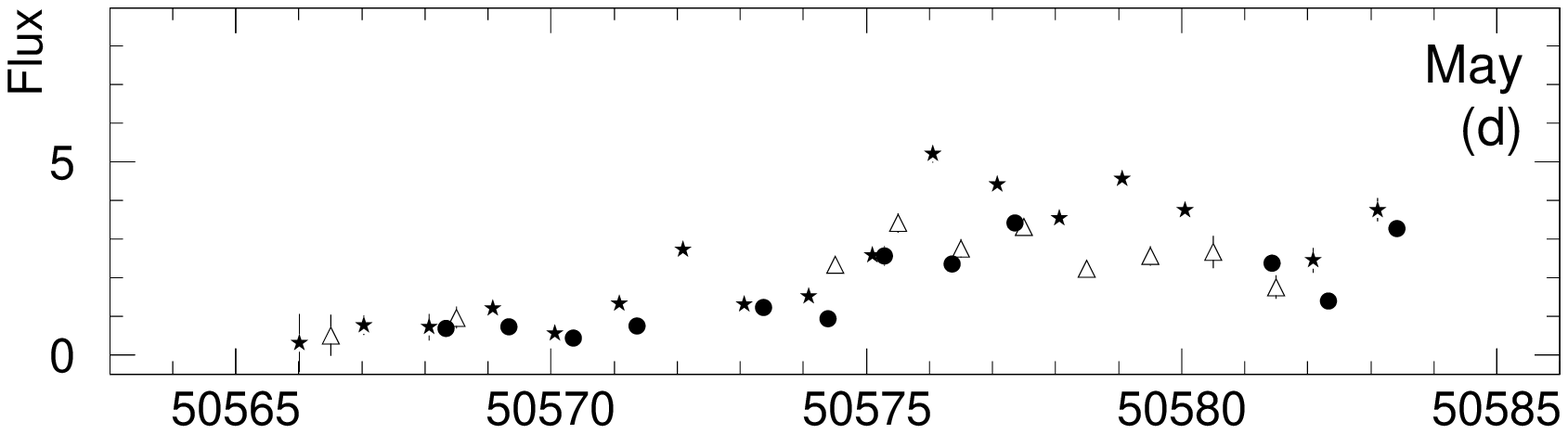,width=3.1in}}
\centerline{\epsfig{file=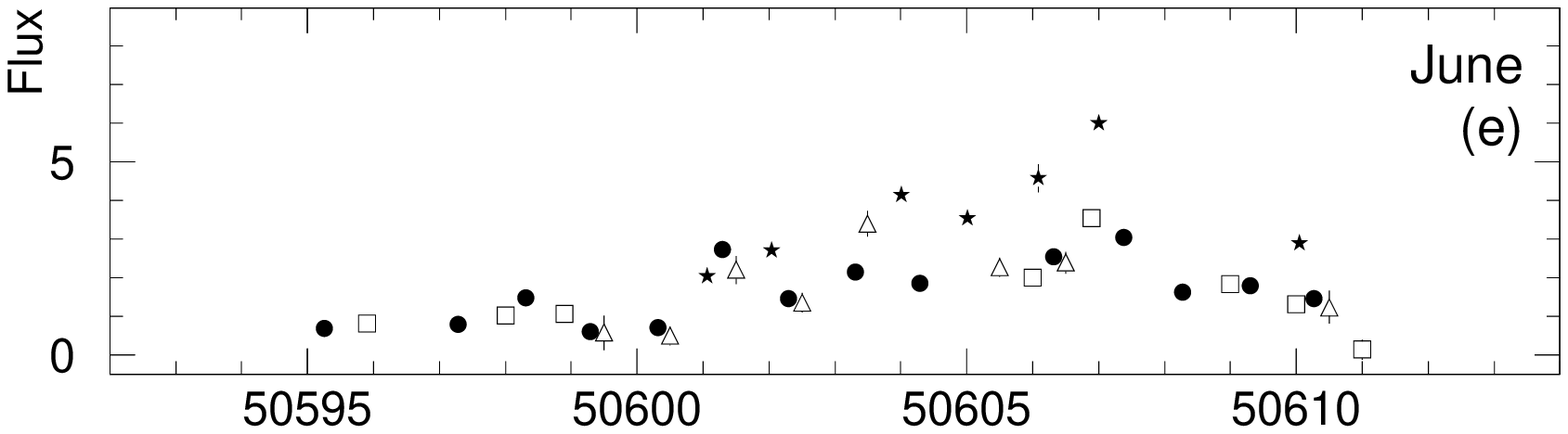,width=3.1in}\hspace*{0.1in}\epsfig{file=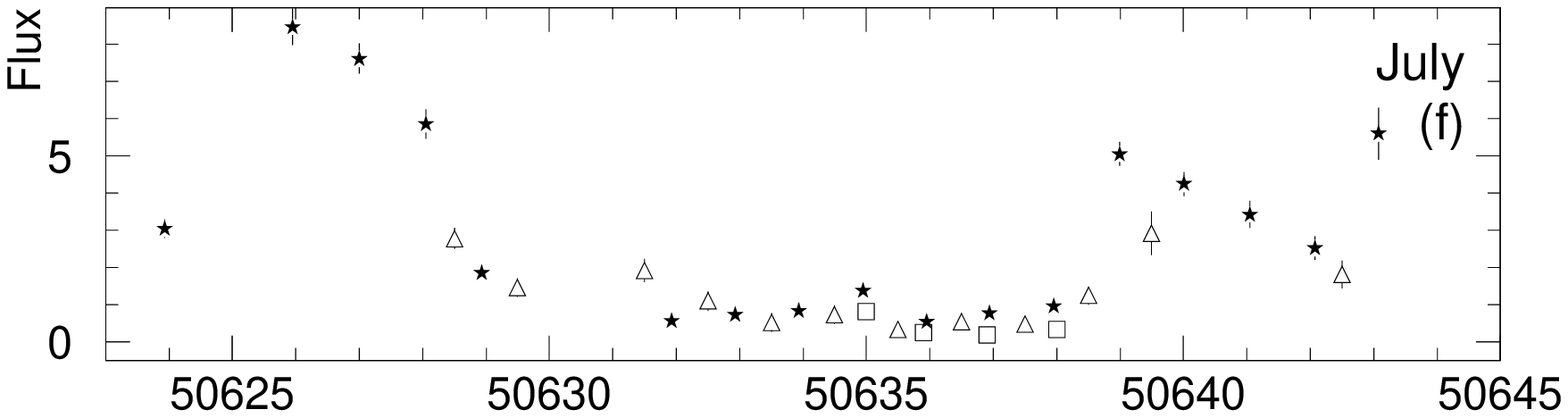,width=3.1in}}
\centerline{\epsfig{file=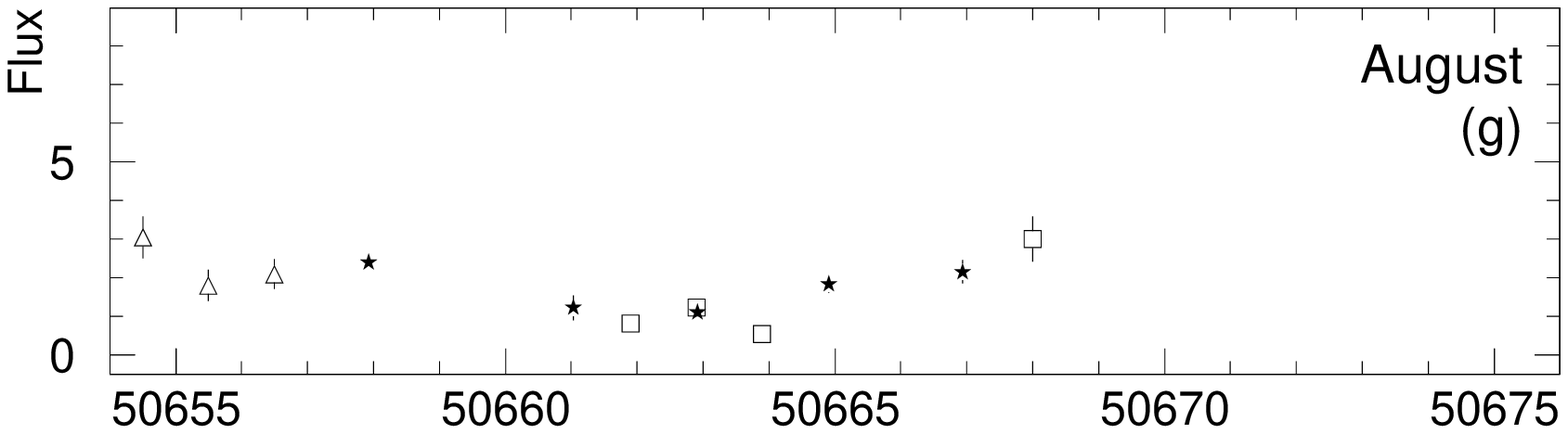,width=3.1in}\hspace*{0.1in}\epsfig{file=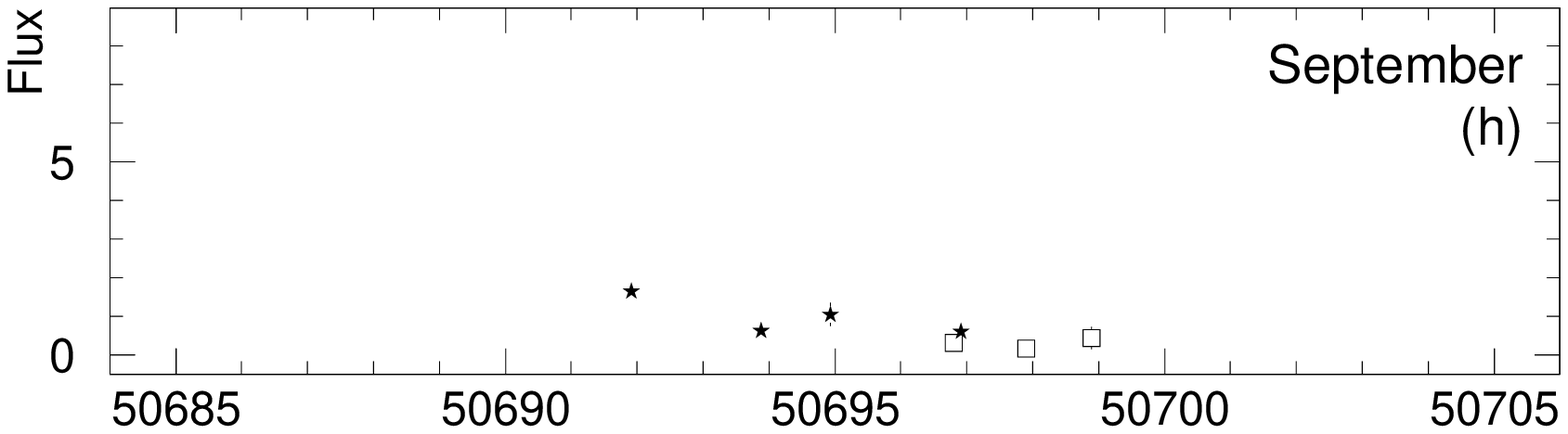,width=3.1in}}
\centerline{\epsfig{file=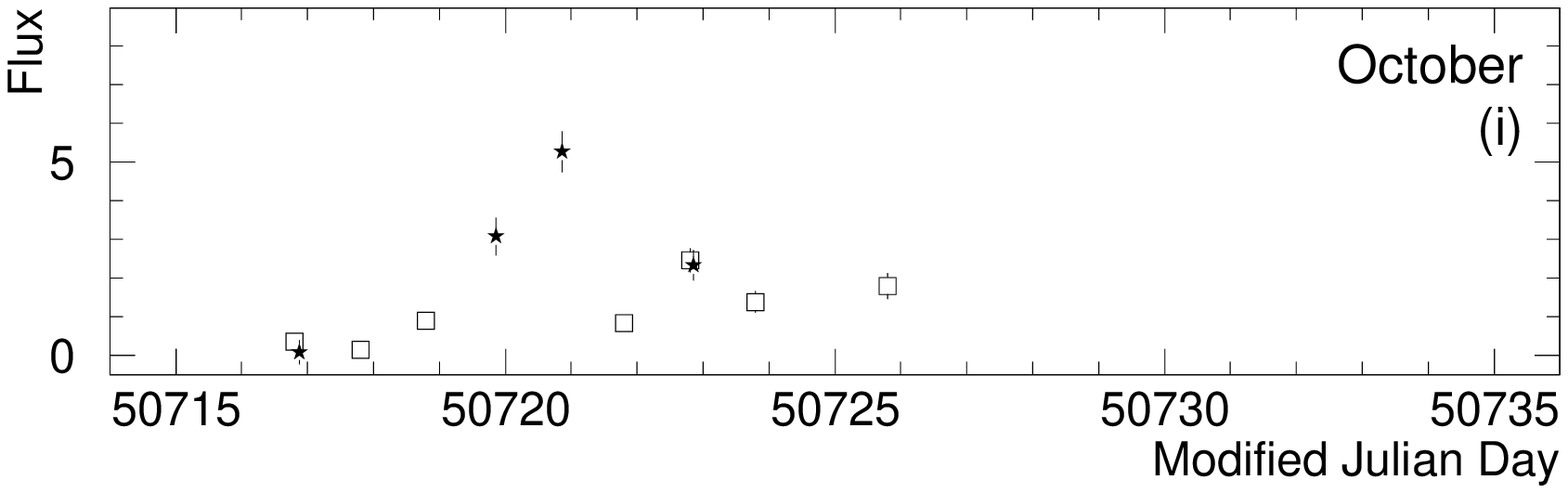,width=3.1in}}
\caption{Very high energy $\gamma$-ray observations of Mrk\,501 with
the Whipple (filled circles, E$>$350\,GeV), CAT (open squares,
E$>$250\,GeV), HEGRA (filled stars, E$>$1\,TeV), and Telescope Array
(open triangles, E$>$600\,GeV) telescopes between 1997 February and
October.  Fluxes are shown in units of the Crab Nebula flux for the
energy threshold of each telescope.  The numbers on the horizontal
axes for each plot indicate the Modified Julian Days during that
month.  Data are from Quinn et al. (1999), Aharonian et al. (1999a),
Djannati-Atai et al. (1999), and Hayashida et al. (1998).
\label{m5_1997_lc}
}
\end{figure*}

Perhaps the most important aspect of the observations of Mrk\,501 in
1997 was that, for the first time, Cherenkov telescopes other than the
Whipple telescope consistently detected a significant excess from
Mrk\,501 on a nightly time-scale.  This permitted more complete VHE
light-curves to be obtained (which it is
expected will eventually lead to a better
understanding of the VHE emission from blazars) and also provided
confirmation that different VHE telescopes could obtain consistent
results from a variable source (see Figure~\ref{m5_1997_lc}).

In addition to the establishment of the flaring itself, the Telescope
Array Collaboration performed a periodicity search with their VHE
observations of Mrk\,501 in 1997 (\cite{Hayashida98}).  They show
evidence for a quasi-periodic signal in the data which has a period of
approximately 12.7 days.  This is disturbingly close to half the lunar
cycle.  If real this would be an extraordinary result, given the very
short time-scale of the quasi-periodicity.

The only published results on observations of Mrk\,501 in 1998 are
those of the Whipple Observatory (\cite{Quinn99}).  The average
emission level was approximately 30\% of the Crab flux but there was
considerably more variability in the emission than in 1996 or 1995.
Two distinct, very high flux flares were observed, one with the
highest flux (approximately 5 times the Crab flux) ever observed from
Mrk\,501 with the Whipple telescope.  The monthly average flux was
also variable, with three months showing emission levels similar to
the 1995 flux, approximately 10\% of the Crab Nebula.

A natural question to ask about the variability in Mrk\,501 is whether
the degree of variability seen changes as a function of the mean flux
level.  That is, whether Mrk\,501 is really more variable when its average
flux is higher or whether it is an artifact of the telescopes being more
sensitive to variations when the average flux is higher?  To test
this, Quinn et al. (1999) performed simulations to see if the
day-scale variability observed with the Whipple telescope in 1997
(when the average VHE flux was 1.3 times that of the Crab) would have
been detectable in 1996 and 1995 (when the average VHE flux was 20\% 
and 10\% that of the Crab, respectively) and also tested whether
the month-scale variations in 1997 and 1996 would have been detectable
in 1995.  Their simulations indicate that the day-scale flaring in
1997 would have been detectable in the 1996 data, but not the 1995
data, and the month-scale variations in 1996 would have been
detectable in 1995, while the 1997 month-scale variations would not
have been detectable.  Thus, it appears that the higher state emission
levels have different variability characteristics than the lower
emission levels.

The other unconfirmed sources are, if they are indeed sources, also
variable emitters of VHE $\gamma$-rays.  1ES\,2344+514 was only
detected with high statistical significance on one night, but has
never been detected since, with flux limits of approximately 10\% of
the Crab flux (\cite{Catanese98}; \cite{Aharonian99b}).  PKS\,2155-304
has also been claimed to be variable (\cite{Chadwick99}), although the
statistical probability that the emission is constant is not quoted.
Finally, 3C\,66A must be variable, or else
observations with the Whipple Observatory telescope
(\cite{Kerrick95b}) and the HEGRA telescope (\cite{Aharonian99b})
would have easily detected this object at the flux level quoted by the
Crimean group.

The high flux VHE emission from Mrk\,421 and Mrk\,501 has permitted
detailed spectra to be extracted.  Accurate measurements of the VHE
spectrum are important for a variety of reasons.  First, the shape of
the high energy spectrum is a key input parameter of AGN emission
models, particularly as it relates to the MeV-GeV measurements by
EGRET.  Second, how the spectrum varies with flux, compared to longer
wavelength observations provide further emission model tests.  Third,
spectral features, such as breaks or cut offs, can indicate changes in
the primary particle distribution or absorption of the $\gamma$-rays
via pair-production with low energy photons at the source or in
intergalactic space (see \S\,\ref{ebl}).

\begin{figure*}[t]
\begin{minipage}{3.4in}
\centerline{\epsfig{file=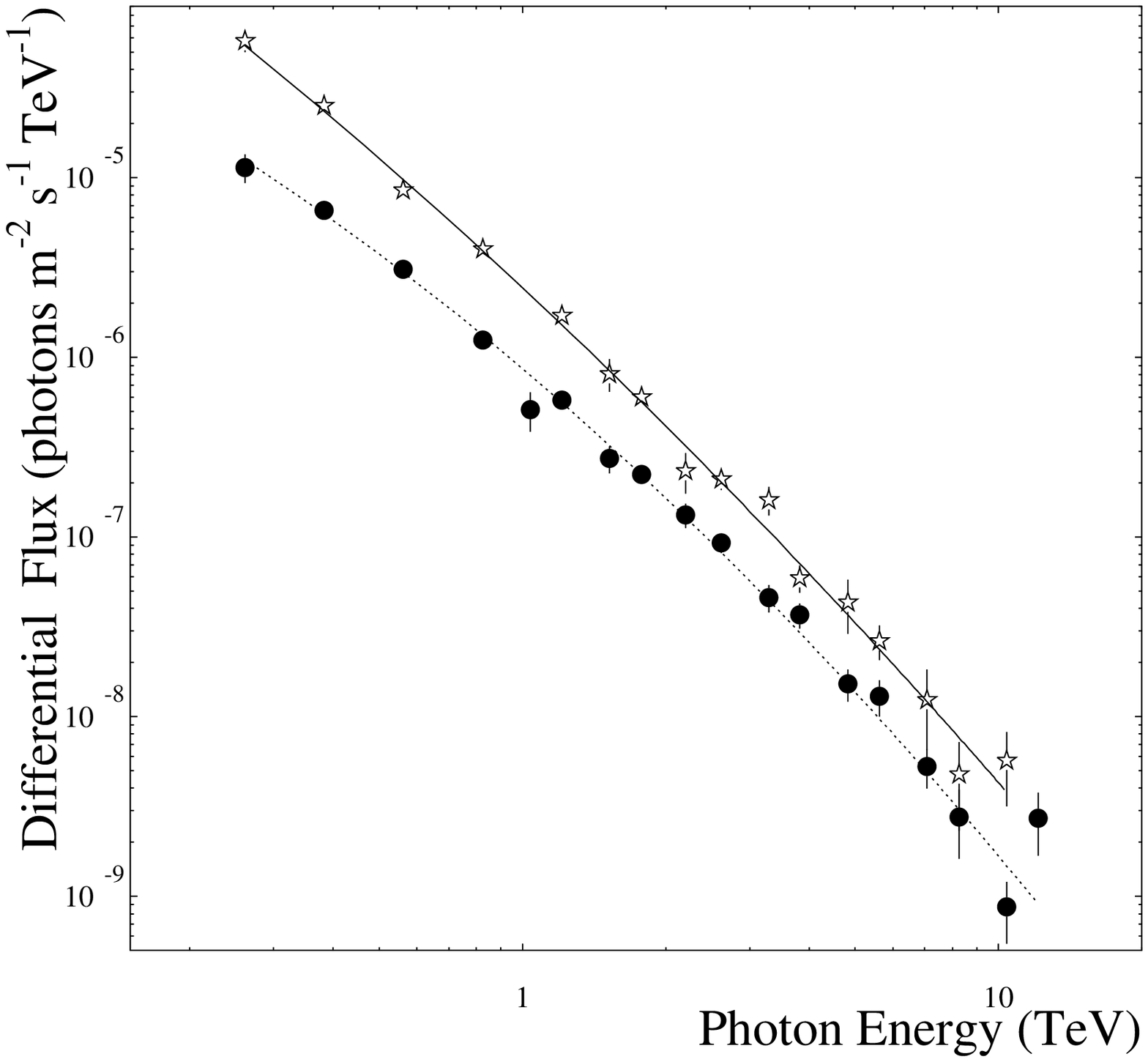,width=3.3in}}
\caption{VHE $\gamma$-ray spectra of Mrk\,421 (open stars) and
Mrk\,501 (filled circles) as measured with the Whipple Observatory
telescope (from \protect\cite{Krennrich99}).  The solid line through
the Mrk\,421 points indicates the best-fit spectrum to these data and
the dashed line through the Mrk\,501 points is the best-fit spectrum
for those data.
\label{m4_m5_whip_spect}
}
\end{minipage}
\hfill
\begin{minipage}{3.4in}
\centerline{\epsfig{file=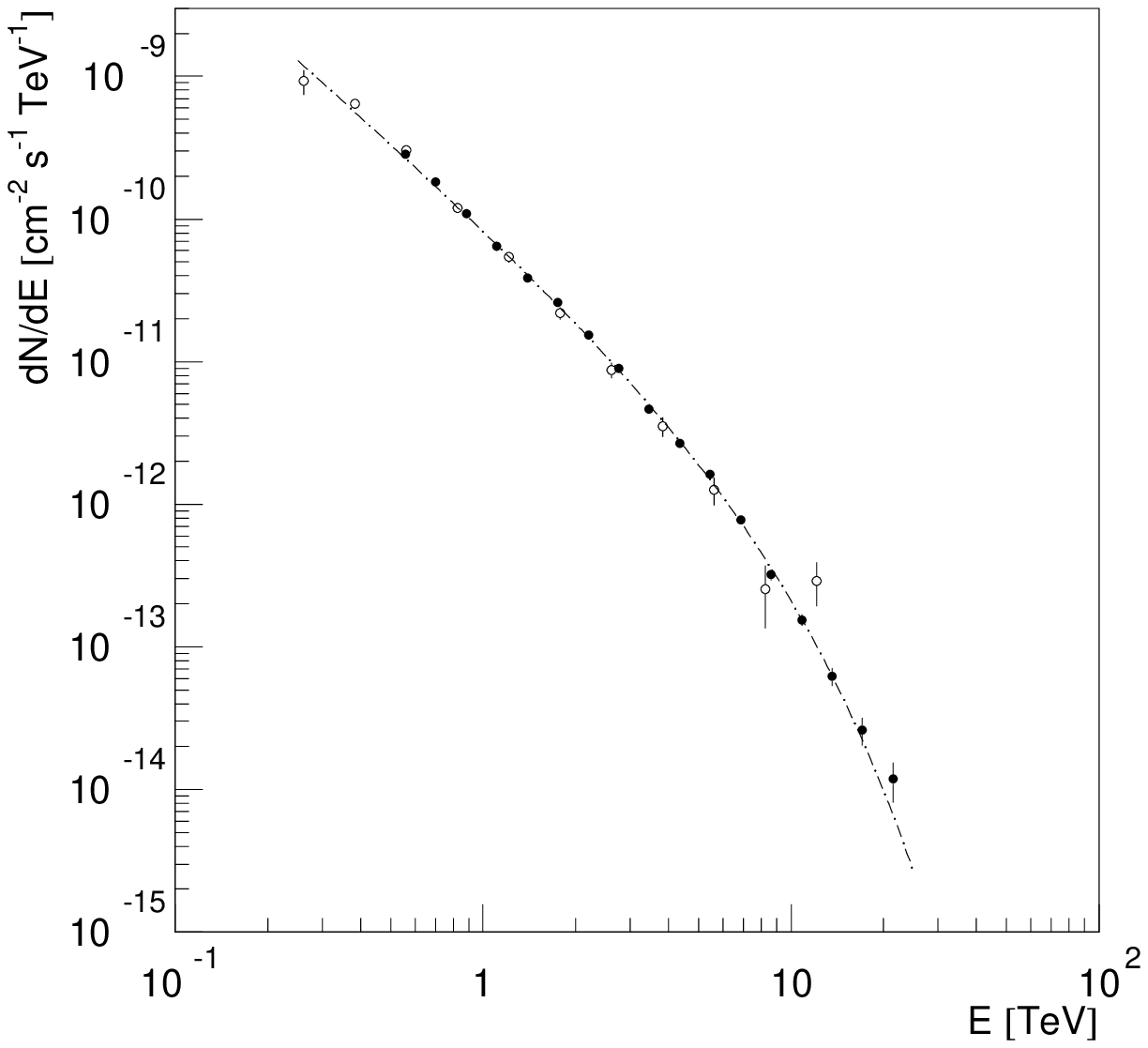,width=3.3in}}
\caption{The energy spectrum of Mrk\,501 as measured by the HEGRA
array (filled circles) and the Whipple telescope (open circles).  The
dashed line indicates the power law plus exponential cut-off spectrum
fit to the HEGRA data.  Figure from Konopelko (1999).
\label{m5_spectra}
}
\end{minipage}
\end{figure*}

For Mrk\,421, the only detailed spectra published at this time come
from observations of high state emission with the Whipple Observatory
telescope (Figure~\ref{m4_m5_whip_spect}; \cite{Zweerink97};
\cite{Krennrich99}).  Analysis of the spectra obtained from
observations of flares on 1996 May 7 and 15 and observations of high
state emission taken at large zenith angles in 1995 June indicate
that, within the statistical uncertainties, the spectra are all
consistent with a simple power law spectrum: ${\rm dN/dE \propto
E^{-2.5}}$ (\cite{Krennrich99}).  When combined, these three data sets
are consistent with a simple power law spectrum for Mrk\,421 of the
form (\cite{Krennrich99})
$${\rm {{dN}\over{dE}} \ (250\,GeV - 10\,TeV) \ \propto E^{-2.54 \pm
0.03_{stat} \pm 0.10_{syst}}}$$
where E is in TeV.

Observations of Mrk\,421 in 1997 and 1998 with the HEGRA system of
Cherenkov telescopes reveal a significantly different spectrum
(\cite{Aharonian99c})
$${\rm {{dN}\over{dE}} \ (500\,GeV - 7\,TeV) \ \propto E^{-3.09 \pm
0.07_{stat} \pm 0.10_{syst}}}$$ than observed with the Whipple
telescope.  The emission level for the HEGRA observations was
approximately 0.5 times the Crab flux, much lower than the fluxes (1
-- 10 times the Crab flux) used in the Whipple observations.  This may
indicate that the spectrum in Mrk\,421 becomes softer with decreasing
flux.  However, HEGRA observations show no evidence of variability
between observations at fluxes above the Crab flux and those between
one-sixth and one-half the Crab flux and the Whipple results show no
variations in spectral index despite using observations spanning a
10-fold range of fluxes.  Further studies may help resolve these
differences.

As with the studies of the variability of its VHE flux, the high state
emission detected from Mrk\,501 in 1997 allowed detailed spectra to be
derived by several experiments, permitting studies of the
time-dependence of the spectra and providing all-important
cross-checks of the methods used to derive energy spectra.  Because of
the rapid variability of the emission, again the normalization of the
spectra are not of fundamental importance, except perhaps in the
context of multi-wavelength studies which are discussed in
\S\,\ref{multi} below.

The most detailed energy spectra published at this time come from
Whipple observations between 250\,GeV and 12\,TeV (\cite{Samuelson98};
\cite{Krennrich99}) and HEGRA data spanning 500\,GeV to 20\,TeV
(\cite{Aharonian99b}).  The Telescope Array Collaboration has also
derived a spectrum over a slightly narrower energy range (600\,GeV to
6.5\,TeV) (\cite{Hayashida98}).  A search for variability in
the spectrum revealed no significant changes in spectrum with flux or
time (\cite{Samuelson99}; \cite{Aharonian99a}), allowing large data
sets to be combined to derive very detailed energy spectra spanning
large ranges in energy.  The spectra derived by Whipple and HEGRA
deviate significantly from a simple power law.  For Whipple, the
$\chi^2$ probability that a power law is consistent with the measured
spectrum is $2.5\times 10^{-7}$.  This is the first significant
deviation from a power law seen in any VHE $\gamma$-ray source and any
blazar at energies above 10\,MeV.  The Whipple spectrum is:
$${{\rm dN}\over{\rm dE}} \propto {\rm E}^{-2.22 \pm 0.04_{\rm
stat} \pm
0.05_{\rm syst} -(0.47 \pm 0.07_{\rm stat}log_{\rm 10}(E))}$$
and the HEGRA spectrum is:
\begin{eqnarray*}
{{\rm dN}\over{\rm dE}} & \propto & {\rm E}^{-1.92 \pm 0.03_{\rm stat}
\pm 0.20_{\rm syst}} \\
 & & *\exp\left[-{{\rm E}\over{6.2 \pm 0.4_{\rm stat}
(-1.5\, +2.9)_{\rm syst}}}\right]
\end{eqnarray*}
where E is in units of TeV.  The form of the curvature term in the
spectra has no physical significance as the energy resolution of the
experiments is not sufficient to resolve particular spectral models.
The Whipple spectral form is simply a polynomial expansion in logE
v. log(dN/dE) space.  The HEGRA form was chosen presumably because
attenuation of the VHE $\gamma$-rays by pair-production with
background IR photons could produce an exponential cut-off.  In fact,
the Whipple and HEGRA data are completely consistent with each other
as shown in Figure~\ref{m5_spectra}.  The Telescope Array
Collaboration derived a spectrum which is well fit by a simple power
law (${\rm dN}/{\rm dE} \propto E^{-2.5 \pm 0.1}$).  The data from
this spectrum are also consistent with the Whipple and HEGRA spectra.

\subsubsection{Multi-wavelength observations}
\label{multi}

Some of the most exciting results on VHE $\gamma$-ray sources have
come through observations where several telescopes operating at
different wavelengths simultaneously monitor the activity in a blazar.
These multi-wavelength campaigns have involved the larger astronomical
community in the study of VHE sources and served the subsidiary
purpose of confirming the source identifications of the VHE
$\gamma$-ray emitting blazars.

\begin{figure*}[t]
\centerline{\epsfig{file=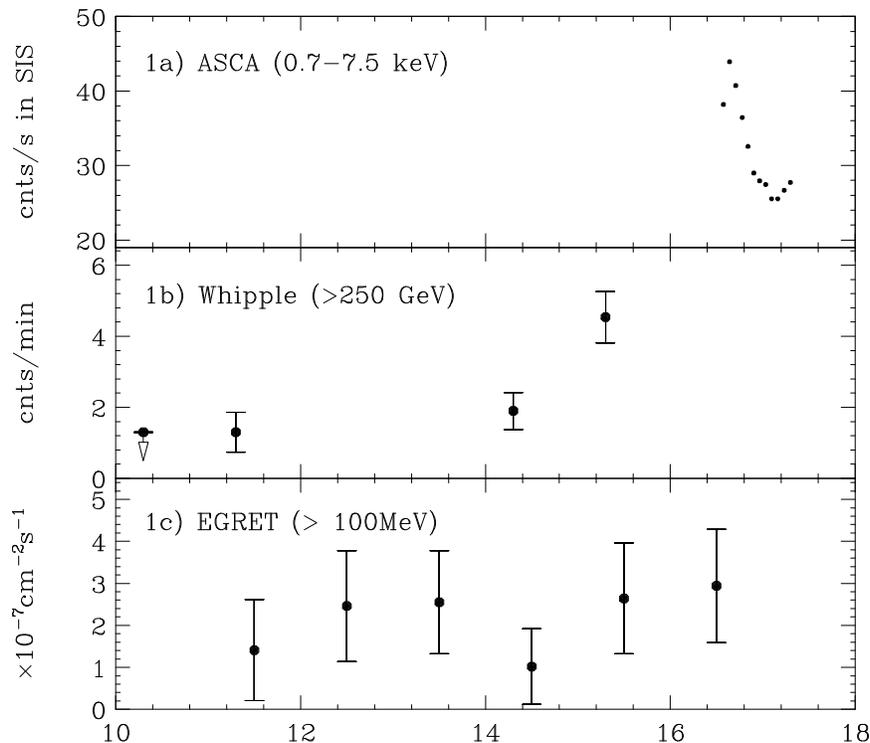,height=4.5in,angle=90.}}
\caption{Lightcurves for observations of Mkn\,421 in
1994 May by {\it ASCA} (top), Whipple (middle), and EGRET
(bottom).  Figure from Takahashi, Madejski, and Kubo (1999).
\label{m4_lc_1994}
}
\end{figure*}

The first evidence of correlated variability between VHE $\gamma$-rays
and lower energy emission came from a multi-wavelength campaign on
Mrk\,421 in 1994 April/May (\cite{Macomb95,Macomb96}).  Observations
were conducted with the Whipple telescope, EGRET, the {\it Advanced
Satellite for Cosmology and Astrophysics} ({\it ASCA}) in X rays, the
International Ultraviolet Explorer (IUE), the United Kingdom Infrared
Telescope (UKIRT), the James Clerk Maxwell Telescope (JCMT) in the mm
waveband, and the University of Michigan 26-m radio telescope (UMRAO)
in the 4.8 -- 14.5\,GHz frequency range.  The VHE $\gamma$-ray, {\it
ASCA}, and EGRET observations are shown in Figure~\ref{m4_lc_1994}.
The VHE observations reveal the rising edge of a flare that developed
over approximately 4 days with the peak flux detected on May 15 (UT)
being approximately 9 times the mean flux measured during that
observing season and approximately 1.4 times the flux of the Crab
Nebula.  Observations had to be halted after May 15 because the phase
of the moon precluded further observations.  Observations taken with
{\it ASCA} on May 16/17 (\cite{Takahashi94}, 1996) indicated a flux
approximately 20 times the quiescent X-ray flux of Mrk\,421, and the
time coincidence between the two observations of unprecedented high
states is the basis for the claim of a correlation in this campaign.
Interestingly, EGRET observations taken between 1997 May 10 and 17 did
not detect the strong day-scale variability seen in VHE $\gamma$-rays.
The average flux during this period was approximately twice the
average flux measured in 1994 April, so there is some evidence of a
higher emission state, but it is not significant enough to claim a
correlation.  The observations at UV, IR, mm, and radio wavelengths
showed no evidence of variability during this period.  Because of the
offset in time of the observations between the VHE $\gamma$-rays and
the X rays, detailed comparisons of the variability in those wavebands
are not possible.

\begin{figure*}[t]
\centerline{\epsfig{file=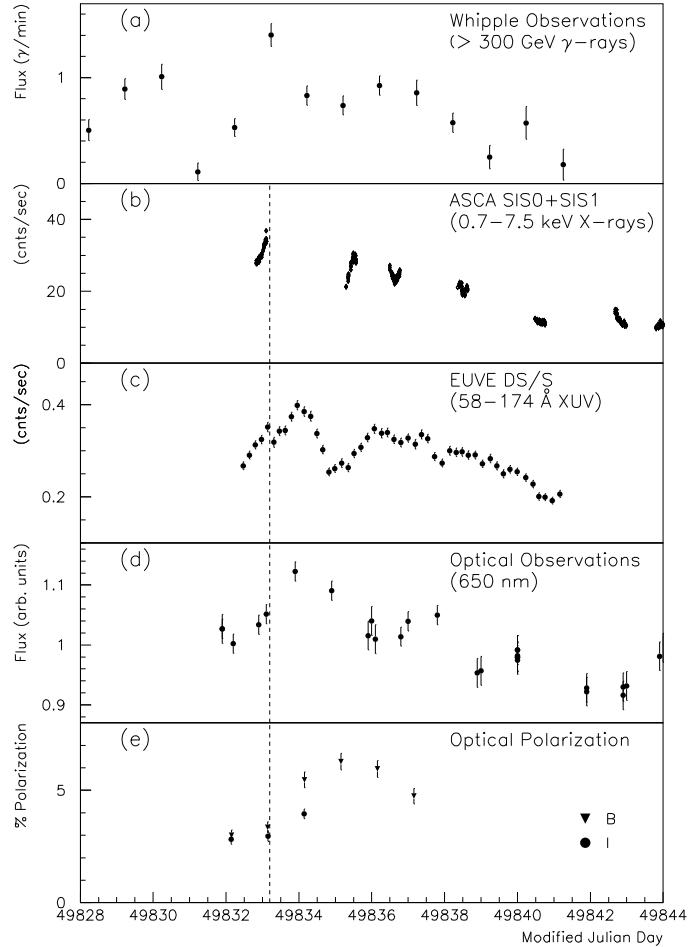,height=5in}}
\caption{(a) Gamma-ray, (b) X-ray, (c) extreme-UV, (d) optical, and 
(e) optical polarization measurements of Mrk\,421 taken 
1995 April -- May.  April 26 corresponds to MJD 49833.  Figure
from Buckley et al. (1996).
\label{m4_multi_95}
}
\end{figure*}

Spurred by this result, another multiwavelength campaign was organized
in 1995 to better measure the multiwavelength properties of Mrk\,421.
This campaign revealed, for the first time, correlations between VHE
$\gamma$-rays and X rays (\cite{Buckley96}).  Observations were
conducted between April 20 and May 5 with the Whipple telescope,
EGRET, {\it ASCA}, the {\it Extreme Ultraviolet Explorer} ({\it
EUVE}), an optical telescope, an optical polarimeter, and UMRAO.
Observations with EGRET did not result in a detection of Mrk\,421.
The 2$\sigma$ flux upper limit for E$>$100\,MeV is $1.2 \times
10^{-7}$ cm$^{-2}$ s$^{-1}$, somewhat below the level detected in
1994.  The light curves for some of these observations are shown in
Figure~\ref{m4_multi_95}.  The optical data have the contribution from
the host galaxy of Mrk\,421 subtracted off.  As in the 1994
multiwavelength campaign, Mrk\,421 underwent a large amplitude flare
in VHE $\gamma$-rays during the observations.  The flare is also
clearly seen in the {\it ASCA} and {\it EUVE} observations.  There is
some evidence for correlated variability in the optical flux and
polarization, but the statistics are not good enough to be confident
about claiming such an association.  The X rays and VHE $\gamma$-rays
appear to vary together, limited to the one day resolution of the VHE
observations, and the amplitude of the flaring is similar, $\sim$400\%
difference between the peak flux and that at the end of the
observations.  The EUVE and optical data (assuming it also is
correlated) are consistent with the flare being delayed by
approximately 1 day relative to the X rays and VHE $\gamma$-rays.  The
amplitude of the flare also decreases with decreasing energy.  The XUV
flux varies by $\sim$200\% during the observations and the optical
flux varies by about 20\%.  The B-band percent polarization varies by
nearly a factor of two in the observations.

The observations of Mrk\,421 in 1994 and 1995 were clearly
undersampled, limiting the conclusions that could be drawn concerning
correlations between wavelengths and emission models.  Two
multiwavelength campaigns organized in 1998 attempted to improve these
measurements through more dense observations in X rays and VHE
$\gamma$-rays.  Improvement in the VHE measurements came about through
longer VHE exposures with individual telescopes, to search for
hour-scale variations, and coordination of VHE observations between
CAT, HEGRA, and Whipple.  Thus, light curves of 12 -- 16 hours could
in principle be achieved, allowing much more detailed measurements of
the VHE emission.  These observations yielded immediate improvements
in the measurements of flaring activity and correlations between VHE
$\gamma$-rays and X rays.

\begin{figure*}[t]
\centerline{\epsfig{file=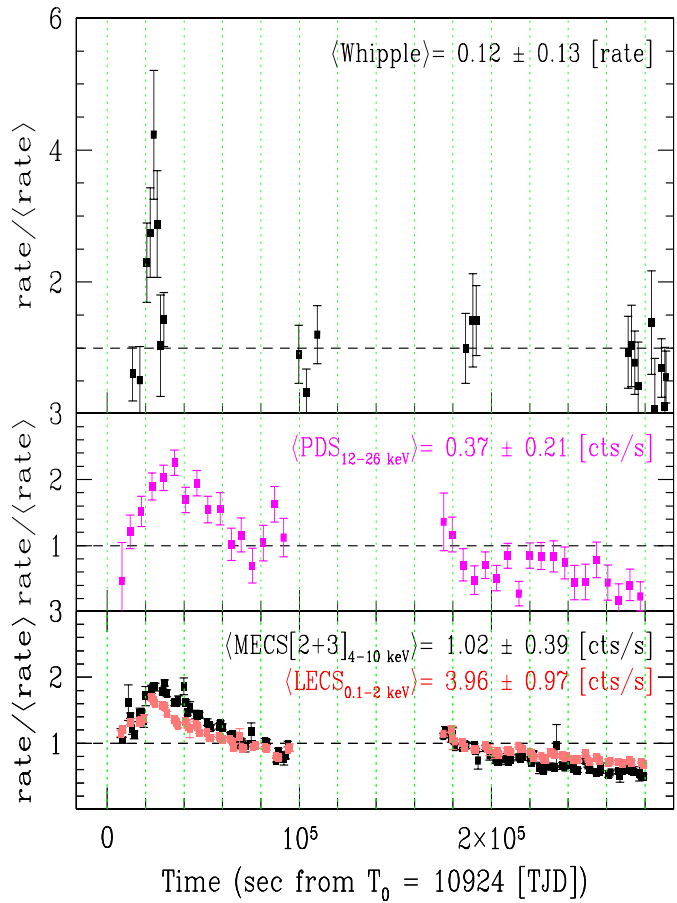,height=5in}}
\caption{Lightcurves for observations of Mrk\,421 in 1998 April by
Whipple and {\it BeppoSAX}.  Whipple observations are for E$>$2\,TeV
and are binned in 28-minute observing segments.  All count rates are
normalized to their respective averages (listed at the top of each
panel) for the observations shown.  Figure from Maraschi et
al. (1999).
\label{m4_1998_sax_lc}
}
\end{figure*}

The first campaign, conducted in late 1998 April, was centered at
X-ray wavelengths on observations with the {\it BeppoSAX} satellite
and established the first hour-scale correlations between X rays and
$\gamma$-rays in a blazar (\cite{Maraschi99}).  The lightcurve for the
observations by {\it BeppoSAX} in three X-ray bands and Whipple above
2\,TeV is shown in Figure~\ref{m4_1998_sax_lc}.  The VHE threshold
here is higher than typical of Whipple data because observations were
taken at a wide range of elevations, causing the energy threshold and
sensitivity to vary with time.  These effects were corrected in the
light curve through the use of simulations and analysis cuts to
normalize the collection area and energy threshold.  Thus, the VHE
lightcurve rate variations are intrinsic.  As
Figure~\ref{m4_1998_sax_lc} shows, a flare is clearly detected in
X rays and TeV $\gamma$-rays on the first day of observations.  The
peaks in the lightcurves occur at the same time, within 1 hour, but
the fall off in the X-ray flux is considerably slower than the TeV
$\gamma$-rays.  Also, the TeV $\gamma$-rays have a larger variability
amplitude ($\sim$4-fold ratio between average and peak) than the
X rays ($\sim$2-fold).  Both the faster VHE flux decrease and the
larger amplitude variability have not been seen previously in
Mrk\,421.  These observations provide the first clear indication that
X rays and VHE $\gamma$-rays may not be completely correlated on all
time-scales.

\begin{figure*}[t]
\centerline{\epsfig{file=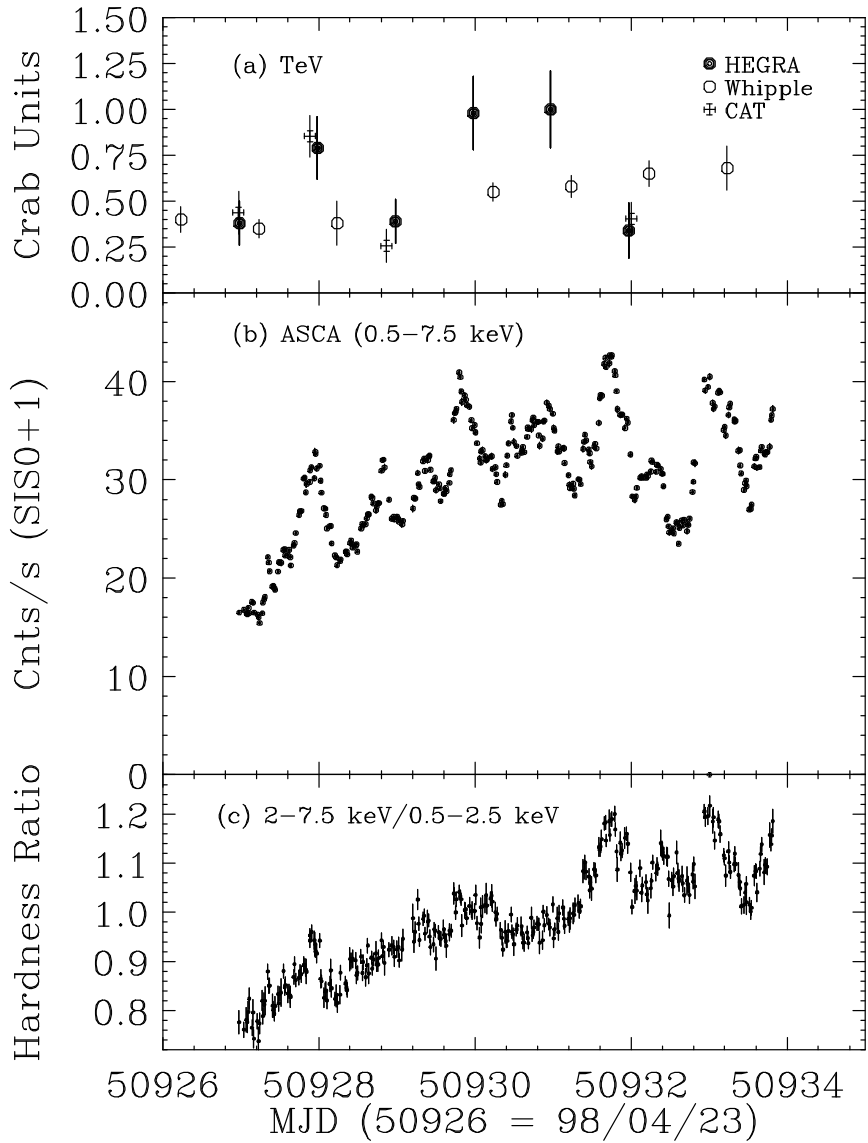,height=4.5in,angle=90.}}
\caption{Lightcurves for observations of Mrk\,421 in 1998 April --
May.  (a) VHE flux observed with HEGRA, Whipple, and CAT, (b) X-ray
flux observed with {\it ASCA}, and (c) X-ray hardness ratios observed
with {\it ASCA} are shown.  Figure from Takahashi et al. (1999).
\label{m4_1998_asca_lc}
}
\end{figure*}

The second multiwavelength campaign started in late 1998 April,
immediately after the observations discussed above, and was centered
around a seven day continuous observation of Mrk\,421 with {\it ASCA}
(\cite{Takahashi99}).  The light curves for the X-ray observations and
VHE observations by Whipple, CAT, and HEGRA are shown in
Figure~\ref{m4_1998_asca_lc}.  Had these observations been conducted
as in 1994 and 1995, with short X-ray exposures and just Whipple
observations, the results would have shown nothing new: similar
variability scale between the X rays and VHE $\gamma$-rays and some
sub-dayscale X-ray variability with amplitude too low to be resolved
with the Whipple observations.  Instead, the X-ray observations reveal
the complete cycle of about 10 flares, the first time this has been
done for Mrk\,421.  Also, these observations seem to confirm the
supposition of Buckley et al. (1996) that the VHE emission from Mrk\,421
is primarily the result of flares, with little steady emission
evident.  Finally, the combination of VHE data from these telescopes
confirms the sub-day-scale correlations seen in the Whipple/{\it
BeppoSAX} observations.  Detailed comparisons of the VHE $\gamma$-ray
and X-ray data will require more sophisticated normalization of the
VHE data (e.g., to common threshold energies) and investigation of
systematics in the measures of variability, but the data hold the
promise of significantly advancing our study of VHE-emitting
$\gamma$-ray blazars.  They also clearly demonstrate the benefits of
operating multiple VHE $\gamma$-ray installations in understanding the
nature of variable sources.

\begin{figure*}[t]
\centerline{\epsfig{file=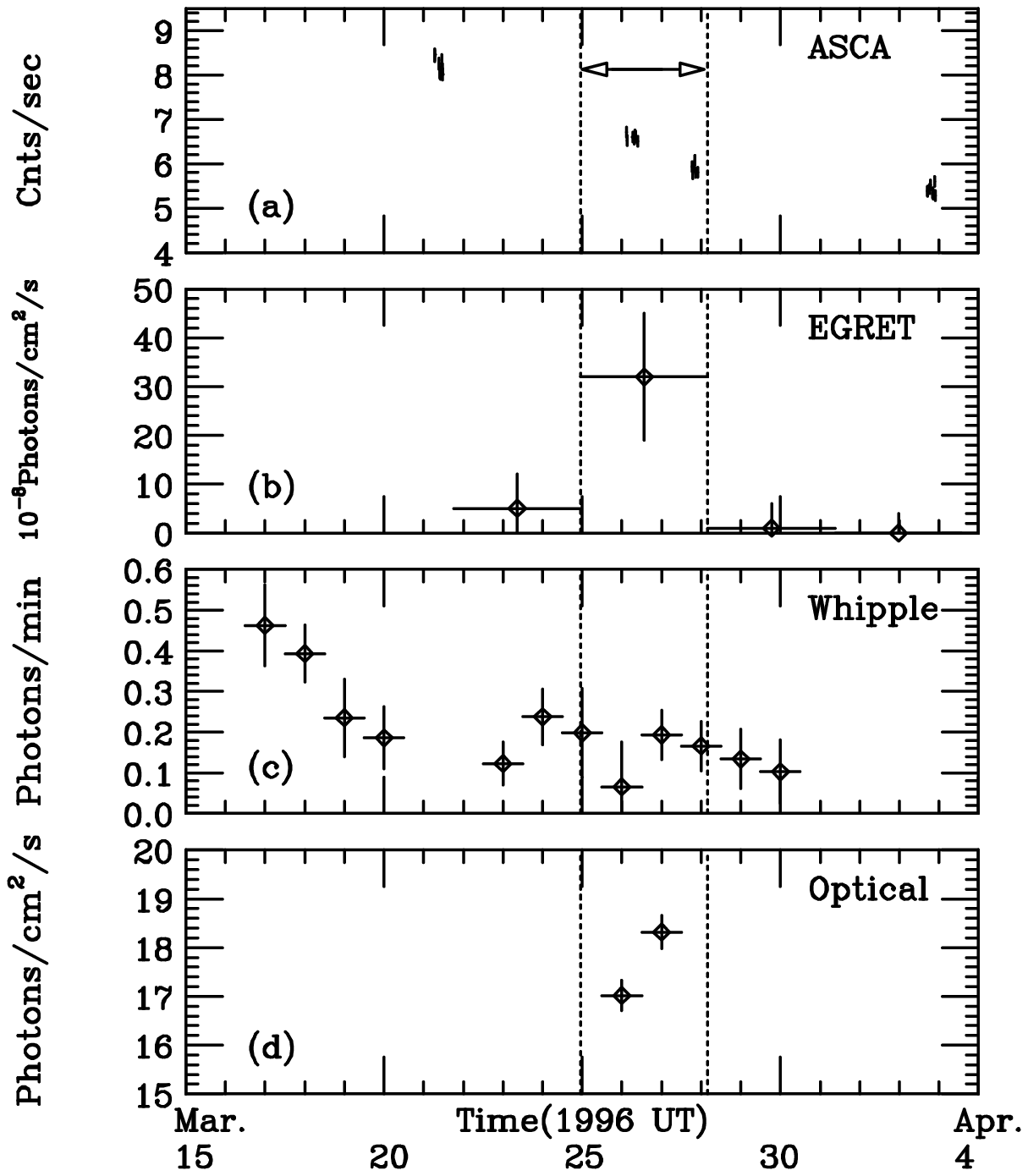,height=4.5in}}
\caption{Lightcurve for observations of Mrk\,501 in 1996 March.
Observations in (a) 0.7 -- 10\,keV X rays ({\it ASCA}), (b)
100\,MeV -- 10\,GeV $\gamma$-rays (EGRET), (c) $>$350\,GeV
$\gamma$-rays (Whipple), and (d) R-band, 650\,nm, optical are shown.
Figure from Kataoka et al. (1999).
\label{m5_lc_1996}
}
\end{figure*}

The first multi-wavelength observations of Mrk\,501 which included VHE
observations were conducted in 1996 (\cite{Kataoka99}).  The
observations were conducted with the Whipple telescope, EGRET, {\it
ASCA}, and an optical telescope.  The light curve for these
observations is shown in Figure~\ref{m5_lc_1996}.  The observations
were too undersampled, or insensitive in the case of EGRET, to clearly
establish any correlations, but these observations have two very
important results.  First, follow up observations in 1996 May
established the first detection of Mrk\,501 by EGRET, with a marginal
significance of 4.0$\sigma$ above 100\,MeV but a significance of
5.2$\sigma$ above 500\,MeV indicating a hard photon spectrum
(\cite{Kataoka99}).  The claim of a 3.5\,$\sigma$ detection by EGRET
during a small part of the multiwavelength campaign (the region
between the dashed lines in Figure~\ref{m5_lc_1996}) seems
speculative, given the lack of any increased activity in other
wavebands during that period.  Second, the observations established a
baseline spectral energy distribution for Mrk\,501 during a relatively
low emission state which could be compared to observations of the high
state emission in 1997 (see below for discussion of spectral energy
distributions).

Multiwavelength observations of Mrk\,501 during its high emission
state in 1997 revealed, for the first time, clear correlations between
its VHE $\gamma$-ray and X-ray emission (\cite{Catanese97c}).
Observations were conducted with Whipple (nightly, from April 7-19),
EGRET and Oriented Scintillation Spectrometer Experiment (OSSE) on
{\it CGRO} (April 9 -- 15), {\it BeppoSAX} (April 7, 11, 16), {\it
RXTE} (twice nightly April 3 -- 16), and the Whipple Observatory
1.2-m optical telescope (nightly, April 7 -- 15).  The optical and
X-ray observations were serendipitously scheduled at this time, and
the {\it CGRO} observations were a public target of opportunity
observation initiated in response to the high VHE emission state.

\begin{figure*}[t]
\centerline{\epsfig{file=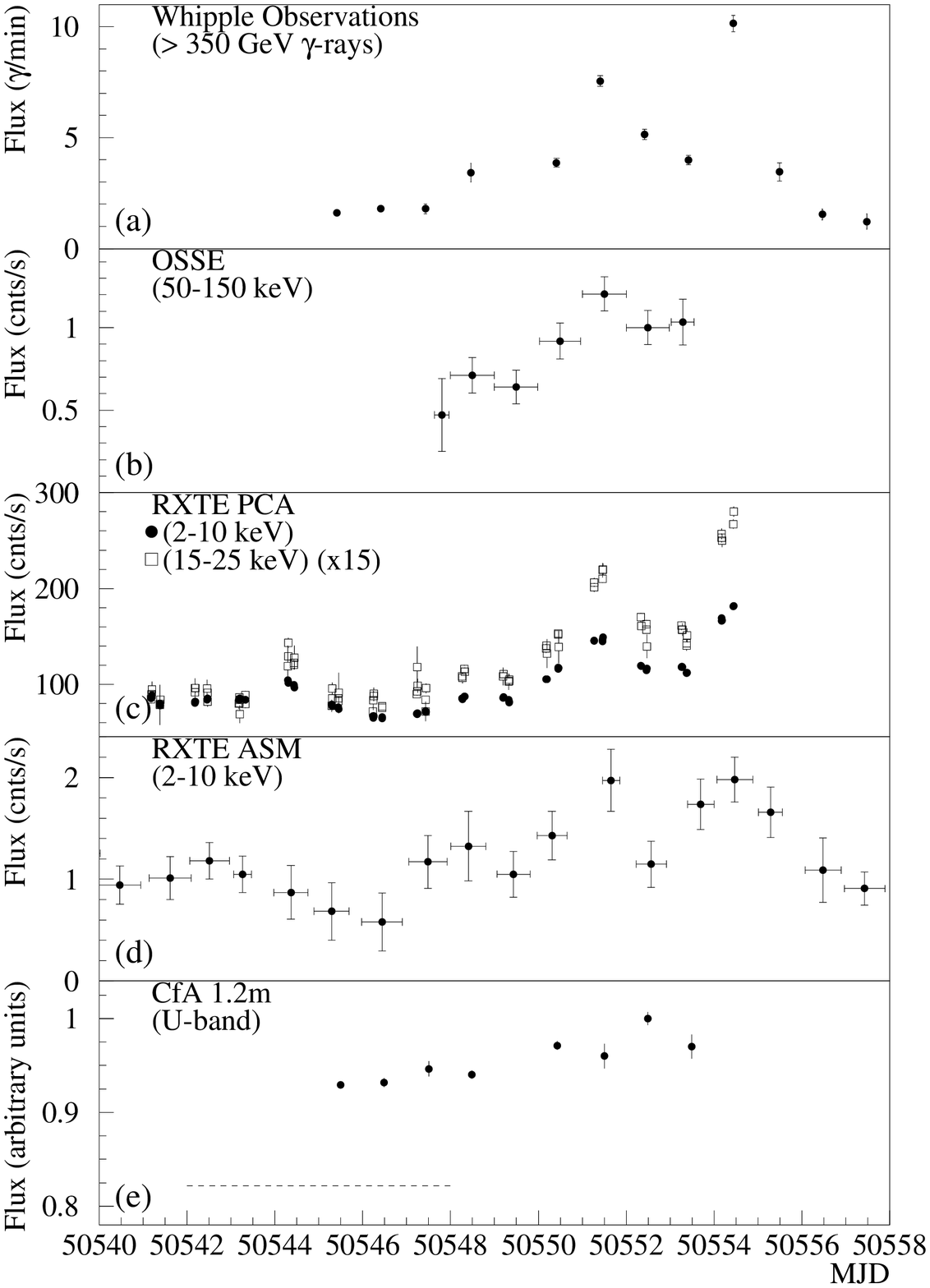,height=5.5in}}
\caption{(a) VHE $\gamma$-ray, (b) OSSE 50 -- 150\,keV, 
(c) {\it RXTE} 2 -- 10\,keV and 15 -- 25\,keV, (d) {\it RXTE} All-Sky
Monitor 2 -- 10\,keV, and (e) U-band optical light curves of Mrk\,501
for the period 1997 April 2 (MJD 50540) to April 20 (MJD 50558).  The
dashed line in (e) indicates the average U-band flux in 1997 March.
Whipple, OSSE, ASM, and optical data are from Catanese et al. (1997)
and {\it RXTE} data are from Catanese (1999).
\label{m5_1997_mw_lc}
}
\end{figure*}

Figure~\ref{m5_1997_mw_lc} shows daily flux levels for the
contemporaneous observations of Mrk\,501.  The average flux level in
the U-band in March is also included in the figure
(Fig.~\ref{m5_1997_mw_lc}e, {\it dashed line}) to indicate the
significant ($\gtrsim$10\%) increase in flux between March and April.
An 11 day rise and fall in flux is evident in the VHE and X-ray
wavebands, with peaks on April 13 and 16.  The 50 -- 150\,keV flux
detected by OSSE also increases between April 9 and 15, with a peak on
April 13.  The optical data may show a correlated rise, but the
variation is small (at most 6\%).  Subtraction of the galaxy light
contribution will increase the amplitude of this variation, but it
should still remain lower than in X rays, given that the R-band
contribution of the galaxy light is $\sim$75\% (\cite{Wurtz96}) and
the U-band contribution should be much less.  EGRET observations
indicated an excess of 1.5$\sigma$, not a significant detection.  The
ratio of the fluxes between April 13 and April 9 are 4.2, 2.6, 2.3,
and 2.1 for the VHE $\gamma$-ray, OSSE, {\it RXTE}, 15 -- 25\,keV, and
{\it RXTE} 2 -- 10\,keV emission, respectively.

The results of this campaign show that for Mrk\,501, like Mrk\,421,
the VHE $\gamma$-rays and the soft X rays vary together and the
variability in the synchrotron emission increases with increasing
energy.  However, OSSE has never detected Mrk\,421 despite several
observations (\cite{McNaron95}), while the Mrk\,501 detection had the
highest 50 -- 150\,keV flux ever detected by OSSE from a blazar.  A
likely explanation of the OSSE detection is that the synchrotron
emission in Mrk\,501 extends to 100\,keV, compared with the
$\sim$1\,keV cutoff seen in Mrk\,421.  This explanation was first
confirmed by the observations with {\it BeppoSAX} (\cite{Pian98}).  In
addition, the day scale variations for Mrk\,501 are larger in
$\gamma$-rays than in X rays, unlike Mrk\,421.  So, despite the
similarity of Mrk\,421 and Mrk\,501 in some respects, these
multi-wavelength campaigns are beginning to reveal differences in the
two objects.

In these short multi-wavelength campaigns, there appears to be a
correlation between the X rays and VHE $\gamma$-rays.  A natural
question to ask is whether this is always true, or only during certain
situations.  An attempt to answer this question has been made by the
HEGRA collaboration by comparing their observations of Mrk\,501 above
500\,GeV to those by the {\it RXTE} All-Sky Monitor, measuring 2 --
12\,keV photons (\cite{Aharonian99a}).  A cross-correlation analysis
of the daily average flux measured by the All-Sky Monitor with the
daily average flux measured by HEGRA reveals a peak in the correlation
function at $\Delta t = 0 \pm 1$ day.  However, the peak in the
cross-correlation function is only $\sim$0.4 and the significance of
the peak is only 2$\sigma$ -- 3$\sigma$.  Whether this indicates that
the X-ray/TeV correlation is {\it not} present is unclear because the
ASM data have large statistical and significant systematic
uncertainties for day-scale measurements of this relatively dim X-ray
source (i.e., compared to the X-ray binaries the ASM was designed to
monitor).  Also, because HEGRA sits on the falling edge of the high
energy spectrum and the ASM sits (for Mrk\,501 in 1997) on the rising
edge of the synchrotron spectrum, it is possible that the emission
detected by these two instruments will not be completely correlated,
particularly for day-scale variations.  Comparison of longer-term
variability between the measurements might help resolve such issues.

\begin{figure*}[t]
\centerline{\epsfig{file=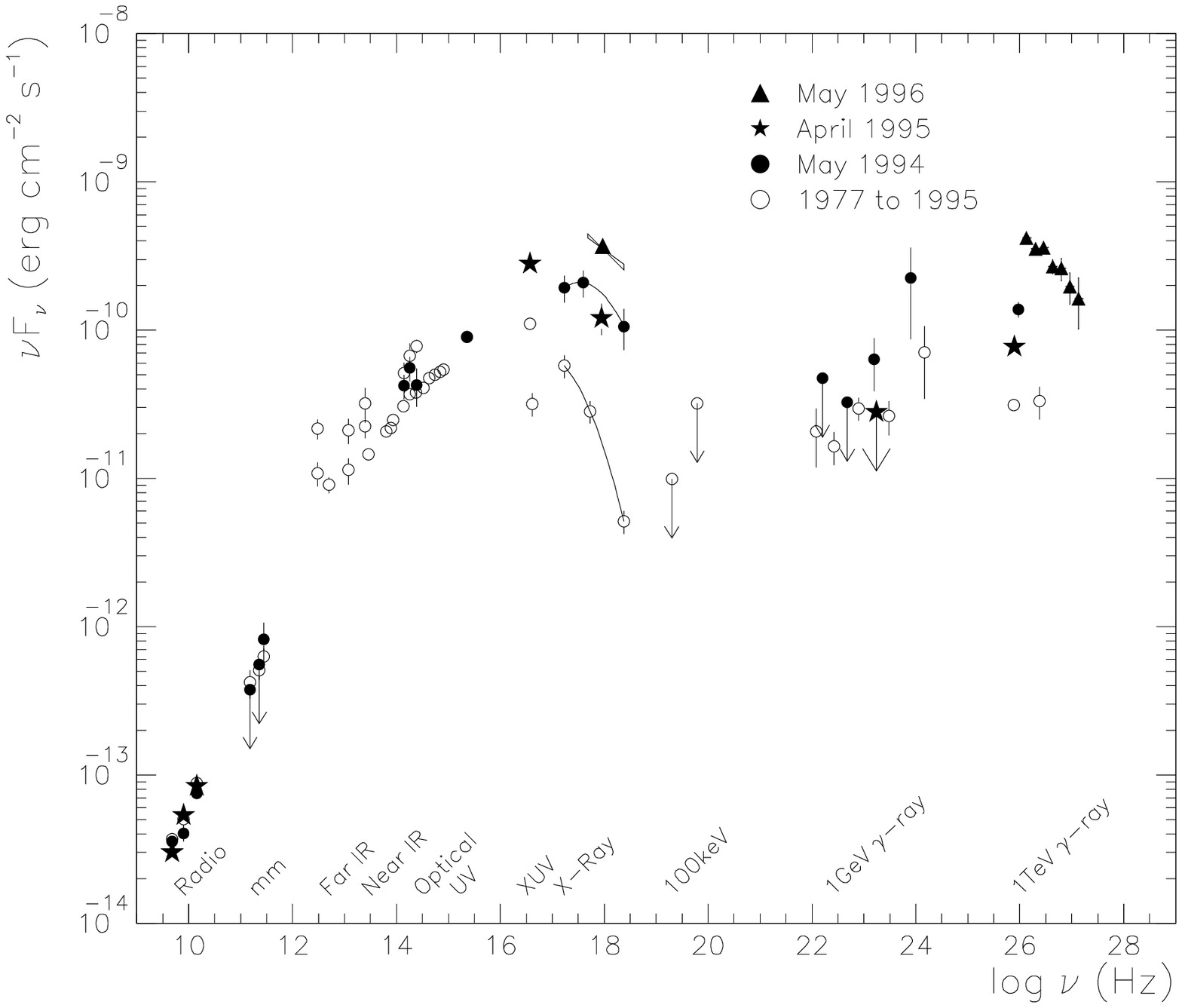,height=2.9in}
\hspace*{0.1in}
\epsfig{file=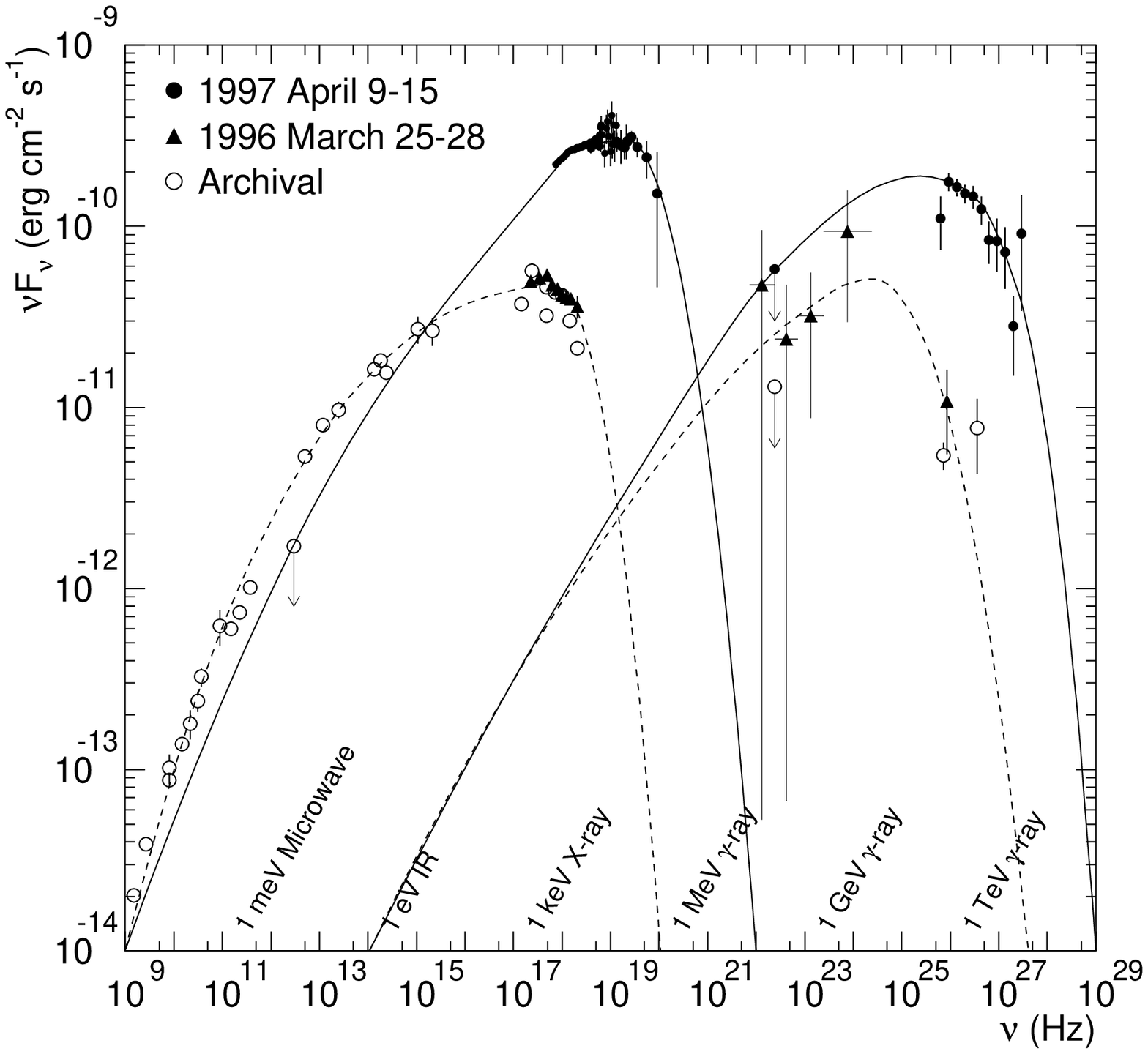,height=2.9in}}
\caption{Left: The spectral energy distribution of Mrk\,421 from
contemporaneous and archival observations.  Dates of the observations
are indicated in the figure.  Figure from Buckley et al. (1997).
Right: The spectral energy distribution of Mrk\,501 from
contemporaneous and archival observations.  Dates of the observations
are indicated in the figure.  Data in the figure come from Kataoka et
al. (1999), Catanese et al. (1997), and Catanese (1999) and references
therein.  The curves in the figure are meant to guide the eye to the
contemporaneously measured points, and do not indicate model fits to
the data nor are they an attempt to elucidate the spectral energy
distribution of Mrk\,501 during these observations.  In both figures,
the archival measurements are approximate averages of the data in the
literature.
\label{seds_fig}
}
\end{figure*}

Figure~\ref{seds_fig} shows the spectral energy distributions (SEDs),
expressed as power per logarithmic bandwidth, for Mrk\,421 and
Mrk\,501 derived from contemporaneous multi-wavelength observations
and an average of non-contemporaneous archival measurements.  Both
have a peak in the synchrotron emission at X-ray frequencies, as is
typical of X-ray selected BL Lac objects, and a high energy peak whose
exact location is unknown but must lie in the 10 -- 250\,GeV range.
Both the synchrotron and high energy peak are similar in power output,
unlike the EGRET-detected flat-spectrum radio quasars which can have
high energy peaks well above the synchrotron peaks (e.g.,
\cite{Montigny95}).  Also, during flaring episodes, the X-ray spectrum
in both objects tends to harden significantly (\cite{Takahashi96},
1999; \cite{Pian98}) while the VHE spectrum is not observed to change.

The SEDs of the two sources do, however, exhibit important
differences.  Most prominent among these is that the combination of
contemporaneous {\it RXTE} and OSSE observations of Mrk\,501 in 1997
clearly confirm the initial measurements of Pian et al. (1998) that
the synchrotron spectrum extended well-beyond the $\sim$1\,keV typical
of X-ray selected BL Lac objects.  They also establish that the peak
power output of the synchrotron emission occurs at $\sim$100\,keV.
This is in contrast to the 1996 observations of Mrk\,501 reported by
Kataoka et al. (1999), where the synchrotron power peak is at
$\sim$2\,keV.  For Mrk\,421, the X-ray spectral peak does shift to
higher energies during flaring activity, but the changes are much
smaller than in Mrk\,501, and the peak was never observed to extend beyond
$\sim$1\,keV.  This peak is followed by a sharp cutoff which produces
a deficit in the OSSE range, preventing the detection of Mrk\,421 by
this instrument.

Whether the shift in the location of the synchrotron peak for Mrk\,501
is also accompanied by a shift in the onset of the $\gamma$-ray
emission to higher energies is not clear.  Any increase in the MeV-GeV
flux in 1997 was not as great as at TeV energies, or EGRET would have
easily detected Mrk\,501.  But, because the sensitivity of EGRET in
1997 was substantially poorer than in 1996, a shift in the onset of
the spectrum or a flux variation that increases with increasing energy
could explain the non-detection by EGRET.  The fact that the VHE
spectrum of Mrk\,501 is harder than that of Mrk\,421 below
$\sim$1\,TeV (see Figure~\ref{m4_m5_whip_spect}) may also indicate
that the high energy peak of Mrk\,501 is shifted to slightly higher
energies.  However, it could also simply indicate a slower fall off in
the progenitor particle spectrum above the peak power output.

A second difference in the spectral energy distributions for these two
objects is that the power output for Mrk\,501 in the VHE range can be
considerably less than in X rays when it is in a low emission state.
In contrast, Mrk\,421 seems to maintain a similar output at X-ray and
$\gamma$-ray energies.  These differences are illustrated in
Table~\ref{sed_ratios} which gives the ratio of contemporaneously
measured fluxes for X rays and $\gamma$-rays for these two objects.
Whether the difference in power output for Mrk\,501 reflects a change
of the energy at which the peak in the high energy spectrum occurs or
something related to the flaring process is not clear, due to the poor
spectral measurements in the low emission states and the lack of
coverage of the peak region of the spectrum.  However, the lack of
spectral variability in $\gamma$-rays argues against a significant
short-term shift of the $\gamma$-ray spectral peak.

\begin{table*}
\caption{Spectral energy ratios for VHE sources \label{sed_ratios}}
\begin{center}
\begin{tabular}{ccrrrc} \hline\hline
Source & Date & \multicolumn{1}{c}{$R_{2\,{\rm keV}}^a$} & 
 \multicolumn{1}{c}{$R_{100\,{\rm keV}}^b$} & 
 \multicolumn{1}{c}{$R_{100\,{\rm MeV}}^c$} & Ref. \\ \hline
Mrk\,421 & 1995 & 2.2$\pm$0.5 & & $<$0.38 & \protect\cite{Buckley96} \\
         & 1996 & 0.81$\pm$0.06 & & & \protect\cite{Buckley97} \\
Mrk\,501 & 1996 & 4.4$\pm$2.1 & & 3.1$\pm$3.4 & \protect\cite{Kataoka99} \\
         & 1997 & 1.2$\pm$0.3 & 1.9$\pm$0.5 & $<$0.36 & 
 \protect\cite{Catanese97c}; \protect\cite{Catanese99} \\ \hline
\multicolumn{6}{l}{$^a$ $\nu$F$_\nu$(2\,keV)/$\nu$F$_\nu$(350\,GeV)} \\
\multicolumn{6}{l}{$^b$ $\nu$F$_\nu$(100\,keV)/$\nu$F$_\nu$(350\,GeV)} \\
\multicolumn{6}{l}{$^c$ $\nu$F$_\nu$(100\,MeV)/$\nu$F$_\nu$(350\,GeV)} \\ 
\end{tabular}
\end{center}
\end{table*}

\subsubsection{Implications of the Very High Energy Observations}
\label{agnimps}

The general properties of the detected extragalactic sources of VHE
$\gamma$-rays are listed in Table~\ref{bllacs-table}.  The three
objects detected by the Whipple Collaboration exhibit some interesting
commonalities.  They are the three closest known BL Lac objects with
declination $>0^\circ$ so their $\gamma$-ray fluxes are the least
attenuated from interaction with background IR radiation.  The
$>$100\,MeV fluxes are near (Mrk 421, Mrk 501) or below
(1ES\,2344+514) the EGRET sensitivity limit, meaning that the
$\gamma$-ray power output does not peak in that energy range as it
does for many of the EGRET-detected AGNs (\cite{Montigny95}).  Thus,
VHE observations already augment the catalog of $\gamma$-ray sources
compiled by space-borne telescopes.  Finally, all three of the
Whipple-detected BL Lac objects are X-ray selected BL Lac objects
(XBLs).  The extension of the synchrotron spectra to X-ray energies in
XBLs implies that they produce high energy electrons, making them good
candidates for VHE emission if the VHE $\gamma$-rays are produced via
inverse Compton (IC) scattering of these same electrons.  EGRET's
tendency to detect more radio-selected BL Lac objects (RBLs) than XBLs
(\cite{Lin97}) also supports this tenet because RBLs would be expected
to have spectra which peak in the MeV-GeV range (\cite{Sikora94};
\cite{Marscher96}).  BL Lac objects in general have been suggested as
better candidates for VHE emission than other blazars because the
absence of optical emission lines in BL Lac objects may indicate less
VHE-absorbing radiation near the emission region (\cite{Dermer94}).

The other two objects detected at VHE energies, PKS\,2155-304 and
3C\,66A, are similar in some respects to the Whipple sources.
PKS\,2155-304 is an XBL, so it fits the IC paradigm for VHE sources.
However, 3C\,66A is classified as an RBL, suggesting that protons produce
the $\gamma$-rays because in IC models, RBLs would not have high
enough energy electrons to produce TeV emission.  In addition, both
PKS\,2155-304 and 3C\,66A are at much higher redshifts than the
Whipple sources, implying quite low IR backgrounds.  Thus,
confirmation of these detections, just as for 1ES\,2344+514, is
essential.

\begin{table*}
\caption{Properties of the VHE BL Lac objects \label{bllacs-table}}
\begin{center}
\begin{tabular}{lrccrrr} \hline\hline
 &  & EGRET flux$^a$ & Average flux & & 
 \multicolumn{1}{c}{$\cal{F}_{\rm X}$ $^a$} & 
 \multicolumn{1}{c}{$\cal{F}_{\rm R}$ $^a$} \\
\multicolumn{1}{c}{Object} & \multicolumn{1}{c}{z} & (E$>$100\,MeV) & 
 (E$>$300\,GeV) &  & \multicolumn{1}{c}{(2\,keV)} & 
 \multicolumn{1}{c}{(5\,GHz)} \\
 &  & (10$^{-7}$cm$^{-2}$s$^{-1}$) & (10$^{-12}$cm$^{-2}$s$^{-1}$) & 
 \multicolumn{1}{c}{$M_v$ $^a$} & \multicolumn{1}{c}{($\mu$Jy)} & 
 \multicolumn{1}{c}{(mJy)} \\ \hline
Mrk\,421 & 0.031 & 1.4$\pm$0.2 & 40 & 14.4 & 3.9 & 720 \\
Mrk\,501 & 0.034 & 3.2$\pm$1.3  & $\geq$8.1 & 14.4 & 3.7 & 1370 \\
1ES\,2344+514$^c$ & 0.044 & $<$0.7 & $\lesssim$8.2 & 15.5 & 1.1 & 220 \\
PKS\,2155-304$^c$ & 0.116 & 3.2$\pm$0.8 & 42 & 13.5 & 5.7 & 310 \\
3C\,66A$^c$ & 0.444 & 2.0$\pm$0.3 & 30$^b$ & 15.5 & 0.6 & 806 \\ \hline
\multicolumn{7}{p{5.6in}}{$^a$Radio, optical, and X-ray data from Perlman et 
 al.\ (1996).  EGRET data from D.J. Thompson (priv. comm.), 
 Mukherjee et al.\ (1997), and Kataoka et al.\ (1999).} \\
\multicolumn{7}{l}{$^b$ $>$1\,TeV flux value.} \\
\multicolumn{7}{l}{$^c$ Unconfirmed as a VHE source.} \\
\end{tabular}
\end{center}
\end{table*}

VHE observations have already significantly affected our
understanding of BL Lac objects.  For example, the rapid variability
indicates either very low accretion rates and photon densities near
the nucleus (Celotti, Fabian \& Rees 1998) or, conversely, requires the
$\gamma$-ray emission region to be located relatively far from the
nucleus to escape the photon fields (\cite{Protheroe97b}).  Also, the
observations have helped resolve the nature of the differences between
RBLs and XBLs.  Based on their smaller numbers and higher
luminosities, Maraschi et al.\ (1986) proposed that RBLs were the same
as XBLs but with jets aligned more closely with our line of sight.
However, the rapid variability and TeV extent of the XBL emission
point to the differences between the two sub-classes being more
fundamental, as originally proposed by Padovani \& Giommi (1994): the
XBLs have higher maximum electron energies and lower intrinsic
luminosities.

Simultaneous measurements of the synchrotron and VHE $\gamma$-ray
spectra also constrain the magnetic field strength ($B$) and Doppler
factor ($\delta$) of the jet.  If the correlation between the VHE
$\gamma$-rays and optical/UV photons observed in 1995 from Mrk 421
indicates both sets of photons are produced in the same region of the
jet, $\delta \gtrsim 5$ is required for the VHE photons to escape
significant pair-production losses (\cite{Buckley96}).  If the SSC
mechanism produces the VHE $\gamma$-rays, $\delta = 15 - 40$ and $B =
0.03 - 0.9$G for Mrk 421 (\cite{Buckley97}; Tavecchio, Maraschi \&
Ghisellini 1998; \cite{Catanese99}) and $\delta \approx 1.5 - 20$ and
$B = 0.08 - 0.2$G for Mrk 501 (\cite{Samuelson98}; \cite{Tavecchio98};
\cite{Hillas99}).  To match the variability time-scales of the
correlated emission, proton models which utilize synchrotron cooling
as the primary means for proton energy losses require magnetic fields
of $B = 30 - 90$G for $\delta \approx 10$ (\cite{Mannheim93};
\cite{Mannheim98}; \cite{Buckley98}).  The Mrk 421 values of $\delta$
and $B$ are extreme for blazars, but they are still within allowable
ranges and are consistent with the extreme variability of Mrk 421.

In addition, the VHE observations have constrained the types of models
that are likely to produce the $\gamma$-ray emission.  For instance,
the correlation of the X-ray and the VHE flares is consistent with IC
models where the same population of electrons radiate the X rays and
$\gamma$-rays.  The absence of flaring at EGRET energies may also
follow in this context (\cite{Macomb95}) because the lower energy
electrons which produce the $\gamma$-rays in the EGRET range radiate
away their energy more slowly than the higher energy electrons which
produce the VHE emission.  The MeV-GeV emission could then be the
superposition of many flare events and would therefore show little or
no short-term variation.

In the mechanism of Sikora et al.\ (1994), which produces
$\gamma$-rays through the Comptonization of external photons, the
external photons must have energies $<$\,0.1\,eV (in the IR band) to
avoid significant attenuation of the VHE $\gamma$-rays by pair
production.  Sikora et al.\ (1994) point out that there is little
direct observational evidence of such an IR component in BL Lac
objects, but the existence of such a field has been predicted as a
product of accretion in AGNs (\cite{Rees82}).

Models which produce the $\gamma$-ray emission from proton progenitors
through $e^+e^-$ cascades originating close to the base of the AGN jet
have great difficulty explaining the TeV emission observed in Mrk\,421
because the high densities of unbeamed photons near the nucleus, such
as the accretion disk or the broad line region, required to initiate
the cascades cause high pair opacities to TeV $\gamma$-rays (Coppi,
Kartje \& K\"onigl 1993).  Such models also predict that the radius at
which the optical depth for $\gamma$-$\gamma$ pair production drops
below unity increases with increasing $\gamma$-ray energy
(\cite{Blandford95}) and therefore the VHE $\gamma$-rays should vary
either later or more slowly than the MeV-GeV $\gamma$-rays.  This is
in contradiction to the observations of Mrk\,421 in both 1994
(\cite{Macomb95}) and 1995 (\cite{Buckley96}).

\subsection{Extragalactic Background Light}
\label{ebl}

In traversing intergalactic distances, $\gamma$-rays may be absorbed 
by  photon-photon pair production ($\gamma + \gamma \rightarrow e^+
+ e^-$) on background photon fields if the center of mass energy of
the photon-photon system exceeds twice the rest energy of the electron 
(\cite{Gould67}).  The cross-section for this process peaks when 
\begin{equation}
{\rm E}_\gamma\epsilon (1 - \cos\theta) \sim 2 (m_ec^2)^2 = 0.52
({\rm MeV})^2
\end{equation}
where E$_\gamma$ is the energy of the $\gamma$-ray, $\epsilon$ is
the
energy of the low energy photon, $\theta$ is the collision angle
between the two photons, $m_e$ is the mass of the electron, and $c$ is
the speed of light in vacuum.  Thus, for photons of energy near
1\,TeV, head-on collisions with photons of $\sim$0.5\,eV have the
highest cross-section, though a broad range of optical-to-IR
wavelengths can be important absorbers because the cross-section for
pair production is rather broad in energy and spectral features in the
extragalactic background density can make certain wavebands more
important than the cross-section alone would indicate. 

The presence of extragalactic background light (EBL) limits the
distance to which VHE $\gamma$-ray telescopes can detect sources.
This has been put forth as an explanation of the lack of detection of
many of the EGRET-detected AGNs (e.g., \cite{Stecker92}), as discussed
above.  The difficulty in understanding the effect of the EBL on the
opacity of the universe to VHE $\gamma$-rays is that not much is known
about the spectrum of the EBL at present, nor how it developed over
time.  Star formation is expected to be a major contributor to the EBL
(e.g., \cite{Madau96}; \cite{Primack99}), with star formation
contributing mainly at short wavelengths (1 -- 15\,$\mu$m) and dust
absorption and re-emission contributing at longer wavelengths (15 --
50\,$\mu$m) .  So, measurements of the EBL spectrum can serve as
important tracers of the history of the formation of stars and
galaxies (\cite{Dwek98}).  Other, more exotic processes, such as
pre-galactic star formation and some dark matter candidates, might
also contribute distinctive features to the EBL (e.g., \cite{Bond86},
1991).  Thus, measurements of the EBL have the potential to provide a
wealth of information about several important topics in astrophysics.

Experiments that attempt to measure the EBL by directly detecting
optical-IR photons, such as the Diffuse Infrared Background Experiment
(DIRBE) on the {\it Cosmic Background Explorer} ({\it COBE}), are
plagued by foreground sources of IR radiation.  Emitted and scattered
light from interplanetary dust, emission from unresolved stellar
components in the Galaxy, and dust emission from the interstellar
medium are all significantly more intense than the EBL and must be
carefully modelled and subtracted to derive estimates of the EBL.
Currently, EBL detections are available only at 140\,$\mu$m and
240\,$\mu$m (\cite{Hauser98}).  Tentative detections at 3.5\,$\mu$m
(\cite{Dwek98a}) and 400 -- 1000\,$\mu$m (\cite{Puget96}) have also
been reported.

Because VHE $\gamma$-rays are attenuated most by optical-IR photons,
measurements of the spectra of AGNs provide an indirect means of
investigating the EBL that is not affected by local sources of IR
radiation (\cite{Gould67}; \cite{Stecker92}).  The signs of EBL
absorption can be cutoffs, but also simple alterations of the spectral
index (e.g., \cite{Stecker99}), depending on the spectral shape of the
EBL and the distance to the source.  Like direct measurements of the
EBL, this technique has difficulties to overcome.  For instance, it
requires some knowledge of or assumptions about the intrinsic spectrum
and flux normalization of the AGNs or the EBL.  Also, the AGNs
themselves produce dense radiation fields which can absorb VHE
$\gamma$-rays at the source and thereby mimic the effects of the
intergalactic EBL attenuation.

\begin{figure*}[t]
\centerline{\epsfig{file=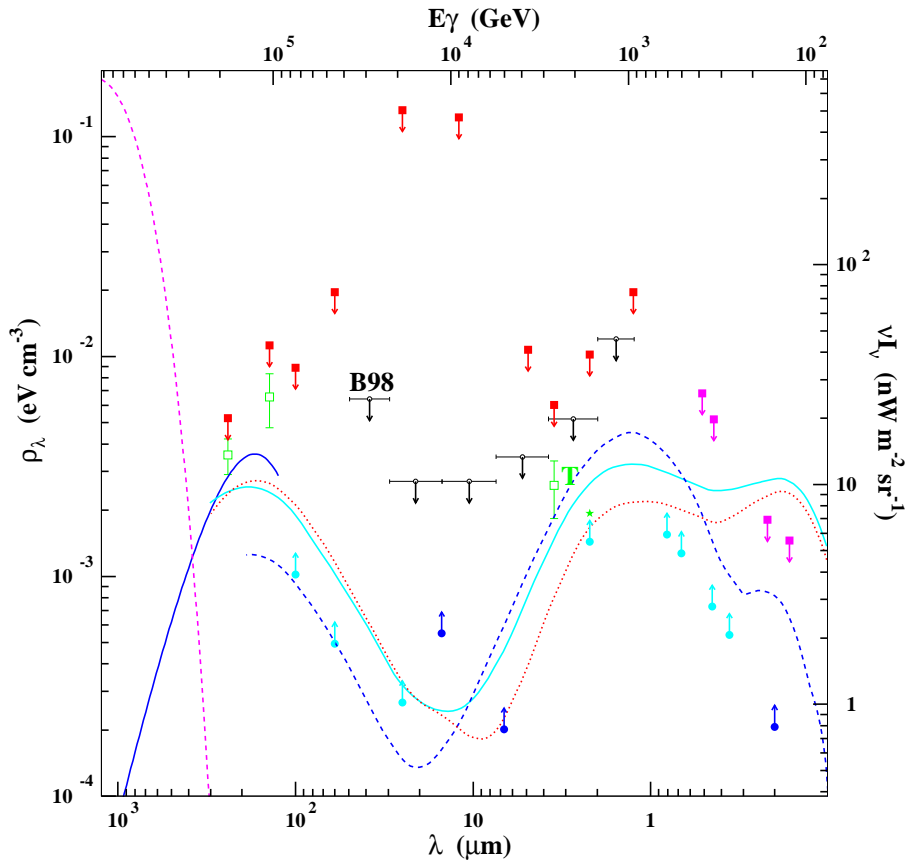,height=4.5in}}
\caption{The diffuse intergalactic infrared background.  E$_\gamma$ is
the energy at which the pair-production cross-section peaks for head on
collisions with photons of wavelength $\lambda$.  Upper limits derived
from VHE $\gamma$-ray spectra are indicated by the horizontal bars
with arrows, marked as B98 (\protect\cite{Biller98}).  Filled
squares are upper limits from various experiments measuring the EBL 
directly (\protect\cite{Hauser98}; \protect\cite{Bowyer91}; 
\cite{Maucherat80}; \protect\cite{Toller83}; \protect\cite{Dube79}).
The open squares at 140\,$\mu$m and 240\,$\mu$m are detections from
DIRBE (\protect\cite{Hauser98}).  The open square marked ``T'' indicates
a tentative detection (\protect\cite{Dwek98a}).  The filled circles are  
lower limits derived from galaxy counts (\protect\cite{Hacking91}; 
\cite{Oliver97}; \cite{Stanev98}; \cite{Pozzetti96}; \cite{Pozzetti98};
\cite{Armand94}).  The solid curve between 90\,$\mu$m and 150\,$\mu$m is
a FIRAS detection (\cite{Fixsen94}).  The dashed line on the left
indicates the 2.7\,K cosmic microwave background radiation.  The three
curves spanning most of the IR wavelengths are different models of 
Primack et al. (1999).  Figure courtesy of V. Vassiliev.
\label{ir-lims}
}
\end{figure*}

\begin{figure*}
\centerline{\epsfig{file=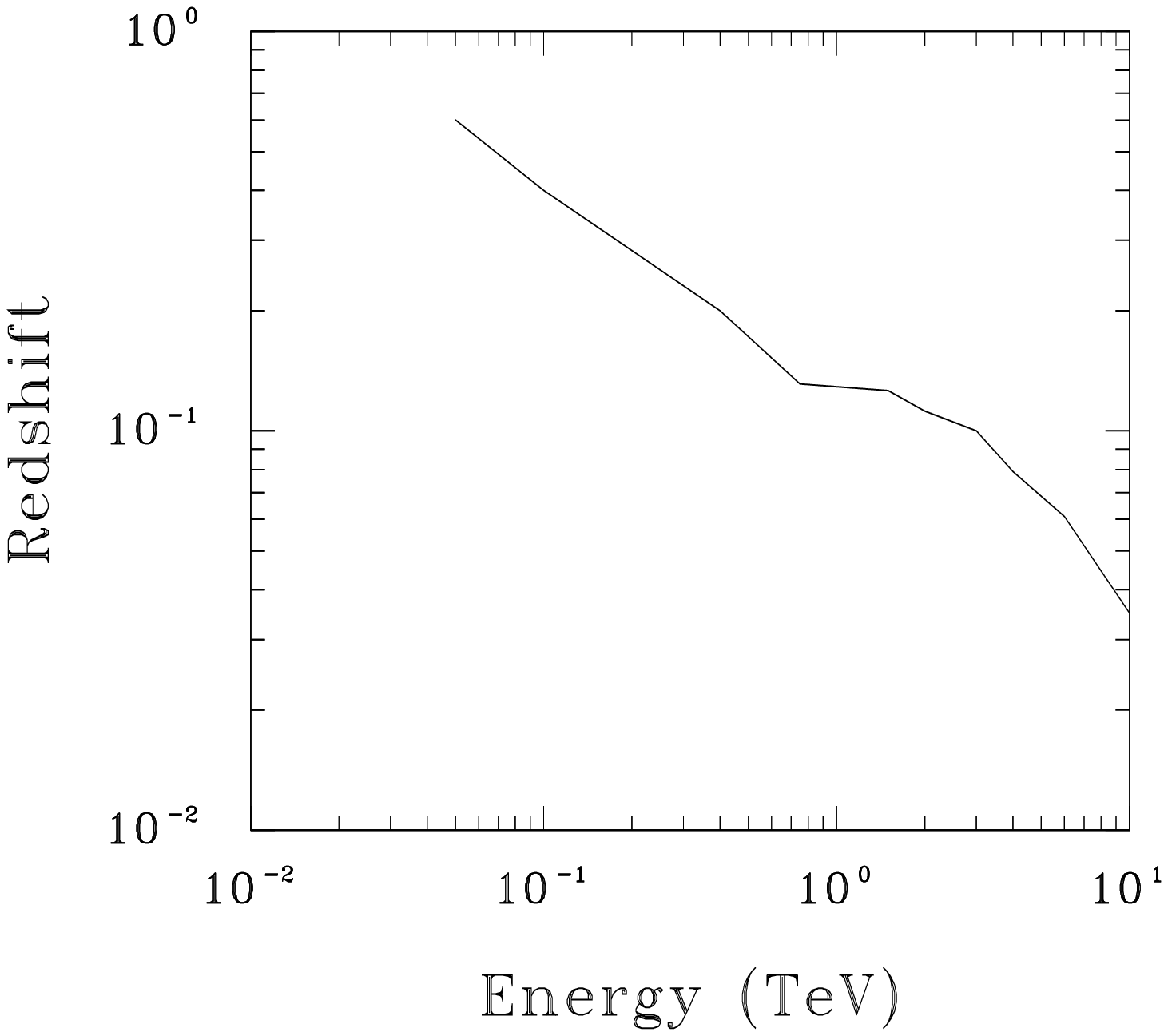,height=3.5in}}
\caption{Lower limit of the redshift at which photons of the energies
shown face an optical depth of 1 due to pair-production with EBa.L
Limits are derived from the upper limits on the density of the EBL of
Biller et al. (1998).  Figure from Biller et al. (1998).
\label{irdist_lim}
}
\end{figure*}

Despite these difficulties, the accurate measurement of VHE spectra
with no obvious spectral cut-offs from just the two confirmed
VHE-emitting AGNs, Mrk\,421 and Mrk\,501 (see \S\,\ref{vhestat}), has
permitted stringent limits to be set on the density of the EBL over a
wide range of wavelengths.  These limits have been derived from two
approaches: (1) assuming a limit to the hardness of the intrinsic
spectrum of the AGNs and deriving limits which assume very little
about the EBL spectrum (e.g., \cite{Biller98}; \cite{Stanev98}) and
(2) assuming some shape for the EBL spectrum, based on theoretical or
phenomenological modelling of the EBL, and adjusting the normalization
of the EBL density to match the measured VHE spectra (e.g.,
\cite{Jager94}; \cite{Stanev98}).  The latter can be more stringent,
but are necessarily more model-dependent.  The limits from these
indirect methods and from the direct measurements of EBL photons are
summarized in Figure~\ref{ir-lims}.  At some wavelengths, the TeV
limits represent a 50-fold improvement over the limits from DIRBE.
These limits are currently well above the predicted density for the
EBL from normal galaxy formation (\cite{Madau96}; \cite{Primack99}),
but they have provided constraints on a variety of more exotic
mechanisms for sources of the EBL (e.g., \cite{Biller98}).  They also
show that EBL attenuation alone cannot explain the lack of detection
of EGRET sources with nearby redshifts at VHE energies, as the optical
depth for pair-production does not reach 1 for the stringent limits of
Biller et al. (1998) until beyond a redshift of $z = 0.1$ (see
Figure~\ref{irdist_lim}).  With the detection or more AGNs,
particularly at higher redshift, and improvements in our understanding
of the emission and absorption processes in AGNs, VHE measurements
have the potential to set very restrictive limits on the EBL density,
and perhaps eventually detect it.

\subsection{Gamma-Ray Bursts}

Although the $\gamma$-ray burst phenomenon is usually associated with
energies of 100 keV to 1 MeV (hard X rays to low energy $\gamma$-rays)
results from EGRET show that there is a component at high energies and
thus the phenomenon has the potential to be observed in the TeV
range. The power spectrum certainly peaks in the lower energy ranges
but the observations at high energies really 
provide the strongest constraints on the emission
models and may ultimately expose the underlying
emission mechanism. The detection of a single photon of energy 18\,GeV
from GRB\,970217, 1.5 hours after the onset of the burst
(\cite{hurley94}), has opened the possibility of delayed emission of
GeV-TeV photons (e.g., \cite{Totani98}; \cite{Bottcher98}).  Although
ACITs have, to date, only presented upper limits (e.g.,
\cite{Connaughton97}), it is possible that in the near future wide
field air shower detectors like MILAGRO or the Tibet Array might 
detect a prompt VHE emission component and rapid slew ACITs might do
the same with greater sensitivity for any delayed emission.

\section{The Future of VHE Astronomy}

It is clear that to fully exploit the potential of ground-based
$\gamma$-ray astronomy the detection techniques must be improved. This
will happen by extending the energy coverage of the technique (with
good energy resolution) and by increasing its flux sensitivity
(improved angular resolution and increased background
rejection). Ideally one would like to do both but in practice there
must be trade-offs.  Reduced energy threshold can be achieved by the
use of larger, but cruder, mirrors and this approach is currently
being exploited using existing arrays of solar heliostats: STACEE
(\cite{chantell98}), CELESTE (\cite{Quebert95}) and Solar-2
(\cite{Tumer99}).  A German-Spanish project (MAGIC) (\cite{barrio98})
to build a 17-m aperture telescope has also been approved. These
projects will achieve thresholds as low as 20-30 GeV where they will
effectively fill the current gap in the $\gamma$-ray spectrum from 20
to 200 GeV. Ultimately, this gap will be covered by GLAST
(\cite{gehrels99}).  Extension to higher energies ($>$10 TeV) can be
achieved by atmospheric Cherenkov telescopes working at large zenith
angles and by particle arrays on very high mountains. An interesting
telescope that has just come on line and will complement these
techniques is the MILAGRO water Cherenkov detector in New Mexico which
will operate 24 hours a day with a large field of view and will have
good sensitivity to $\gamma$-ray bursts and transients
(\cite{Sinnis95}).

\begin{figure*}[t]
\centerline{\epsfig{file=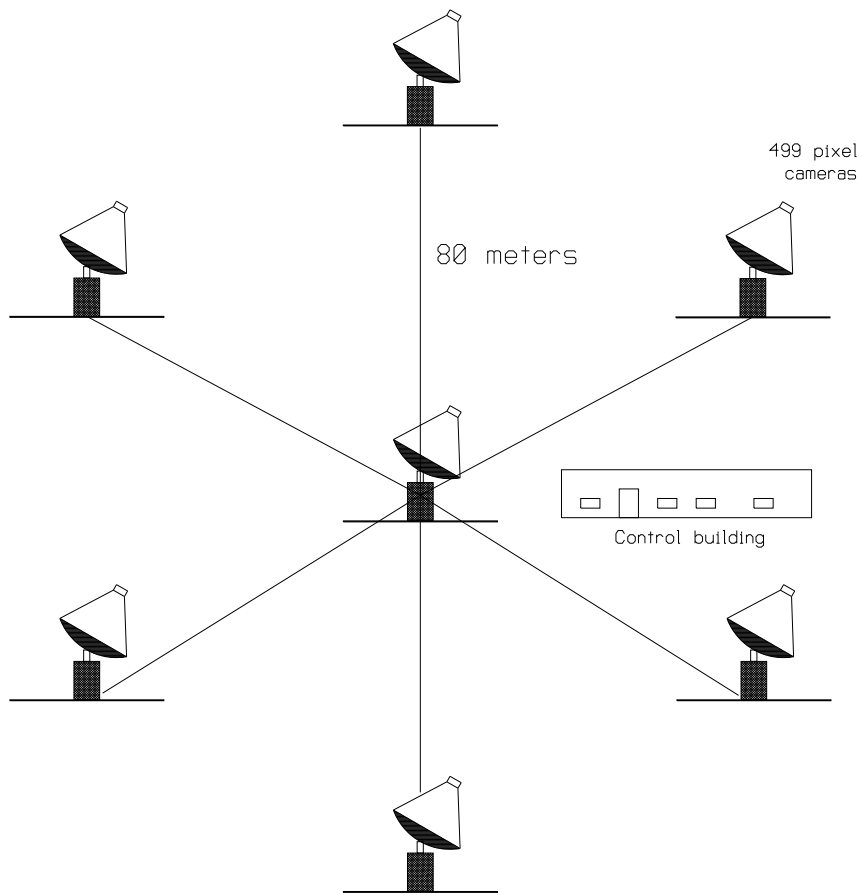,height=4.5in,angle=270.}}
\caption{The proposed arrangement of telescopes in VERITAS.  Figure from
Weekes et al. (1999).
\label{cam-fig}
}
\end{figure*}

VERITAS, with seven 10-m telescopes arranged in a hexagonal pattern
with 80\,m spacing (Figure~\ref{cam-fig}), will aim for the middle
ground between those techniques listed above, with its primary
objective being high sensitivity observations in the 100\,GeV to
10\,TeV range (\cite{weekes99}). It will be located in southern
Arizona and will be a logical
progression from the Whipple telescope. It is hoped to begin
construction in 1999 and to complete the array by 2004.

\begin{figure*}[t]
\centerline{\epsfig{file=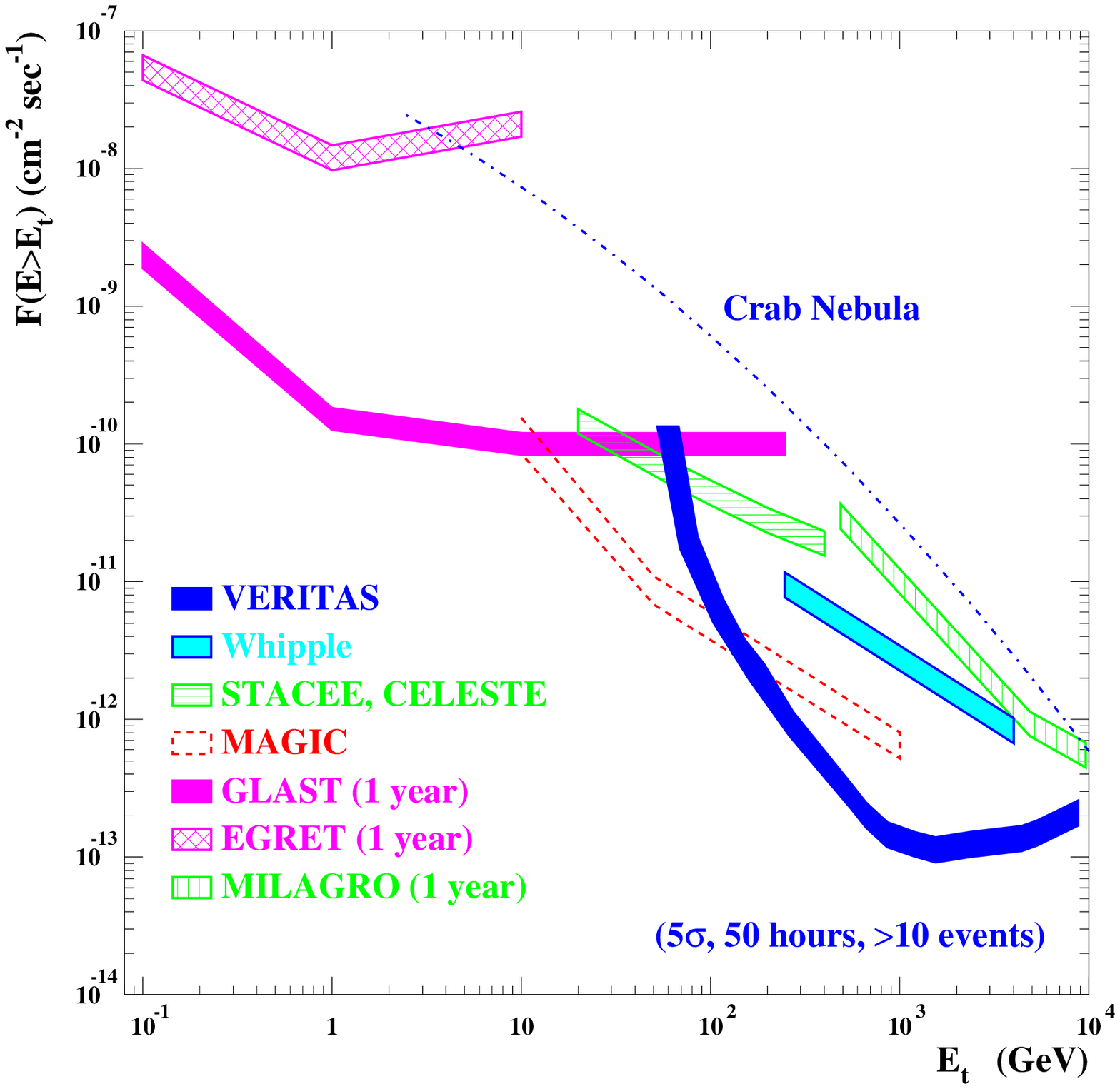,height=4.5in,angle=0.}}
\caption{Comparison of the point source sensitivity of VERITAS to
Whipple (\protect\cite{weekes89}), MAGIC (\protect\cite{barrio98}),
CELESTE/STACEE (\protect\cite{Quebert95}; \protect\cite{chantell98});
HEGRA (\protect\cite{daum97}), GLAST (\protect\cite{gehrels99}), EGRET
(\protect\cite{thompson93}), and MILAGRO
(\protect\cite{Sinnis95}). The sensitivity of MAGIC is based on the
availability of new technologies, e.g., hybrid PMTs, not assumed in
the other experiments. EGRET, GLAST and MILAGRO are wide field
instruments and therefore ideally suited for all sky surveys.  The
turn-up in the VERITAS sensitivity at higher energies is primarily
caused by the requirement that the signal contain at least 10 photons.
\label{main-senscomp-fig}
}
\end{figure*}

The German-French-Italian experiment HESS, initially four and
eventually perhaps sixteen 12m class telescopes in
Namibia (\cite{hofmann97}), and the Japanese Super CANGAROO array,
with four 10-m telescopes in Australia, (T. Kifune, private
communication) will have similar objectives.  In each case, the arrays
will exploit the high sensitivity of ACITs and the high selectivity of
the array approach. The projected sensitivities of MAGIC, HESS,
SuperCANGAROO and VERITAS are somewhat similar and we refer to them
collectively as Next Generation Gamma-Ray Telescopes (NGGRTs).  The
relative flux sensitivities for existing and planned $\gamma$-ray
telescopes as a function of energy are shown in
Figure~\ref{main-senscomp-fig}, where the sensitivities of the wide
field detectors are for one year and the atmospheric Cherenkov
telescopes are for 50 hours.  In all cases, a 5\,$\sigma$ point source
detection is required.

It is apparent from this figure that, on the low energy side
($<$1\,TeV), the NGGRTs will complement the GLAST mission and will
overlap with the solar arrays.  At the highest energies to which they
are sensitive, NGGRTs will overlap with the Tibet Air Shower Array
(\cite{Tibet97}). They will cover the same energy range as MILAGRO but
with greater flux sensitivity. The wide field coverage of MILAGRO will
permit the detection of transient sources which, once detected, can be
studied in more detail by the northern NGGRTs. These same telescopes
will complement the coverage of neutrino sources to be discovered by
AMANDA/ICE CUBE (\cite{Halzen98}) at the South Pole.  Finally, if the
sources of ultra-high energy cosmic rays are localized to a few
degrees by HiRes (\cite{Abu97}) and Auger (\cite{Boratav97}), the
NGGRTs will be powerful instruments for their further localization and
identification.

The recent successes in VHE $\gamma$-ray astronomy 
ensure that in the inevitable interval between the death of EGRET and
the launch of the next generation $\gamma$-ray space telescope, there
will be ongoing activity in GeV-TeV $\gamma$-ray astronomy.  
Observations by GLAST and the NGGRTs in this
energy region will make important contributions to our understanding of
AGNs, supernova remnants, and pulsar and $\gamma$-ray burst studies.
Although the number of TeV sources detected so far is small, the new
and varied phenomena observed indicate that VHE $\gamma$-ray astronomy
is not merely an extension of MeV-GeV $\gamma$-ray astronomy but is a
discipline in its own right. With the advent of new telescopes, the
catalog of VHE $\gamma$-ray sources will dramatically expand 
with detailed time histories and accurate energy
spectra available.

\acknowledgments 

We thank F. Aharonian, A. Djannati-Atai, F. Kajino, J. Kataoka,
M. Punch, T. Takahashi, D. Thompson, and V. Vassiliev for providing some of the data
and figures presented here, and A. Burdett and S. Fegan
for reading the manuscript.  This research is supported by grants from
the U. S.  Department of Energy and NASA.


\end{document}